\documentclass[12pt,final]{article}
\usepackage{amssymb}
\usepackage{a4wide}

\usepackage[only,twltt,tentt]{rawfonts}

\newcommand{\tth}{'014}

\newcommand{\Vb}[1]{`\vb{#1}'}
\newcommand{\vb}[1]{{\def\-{\discretionary{\char\tth}{}{}}\relax%
                     \hyphenchar\twltt=\tth\relax%
                     \hyphenchar\tentt=\tth\relax%
                                           {\tt #1}}}

\newcommand{\grg}{\mbox{$\mbox{GRG}_{\mbox{\tiny EC}}$}}
\newcommand{\Grg}{\mbox{$\mbox{GRG}_{\!\mbox{\scriptsize EC}}$}}
\newcommand{\GRG}%
{\mbox{$\mbox{\bf GRG}_{\hspace{-0.2ex}\mbox{\normalsize\bf EC}}$}}

\newcommand{\firstgap}{0.1} %
\newcommand{\gaps}{0.3} %
\newcommand{\brgaps}{0.4} %

\newcommand{\someskip}{\smallskip}

\newcommand{\Ratio}[2]{{\textstyle{#1\over #2}}}
\newcommand{\half}{\Ratio{1}{2}}

\newcommand{\mbold}[1]{\hbox{\boldmath $#1$}}

\newcommand{\ssqrt}[1]{\hbox{\small$\sqrt{#1}$}}

\newcommand{\Red}{{\sl Reduce\/}}
\newcommand{\Exc}{{\sl Excalc\/}}
\newcommand{\Com}{{\sl COMPACT\/}}

\newcommand{\ie}{\hbox{\em i.\kern -0.12ex e.}}
\newcommand{\apriori}{{\em a priori\/}}

\newcommand{\etc}{{\em etc.}}
\newcommand{\cf}{{\em cf.}}
\newcommand{\Cf}{{\em Cf.}}
\newcommand{\eg}{{\em e.g.\/}}

\newcommand{\vv}{{\em vice versa\/}}
\newcommand{\mod}{{\em modulo\/}}

\newcommand{\re}{\mathop{\mathrm{Re}}}
\newcommand{\im}{\mathop{\mathrm{Im}}}
\newcommand{\imu}{\mathrm{i}}
\newcommand{\dex}{\mathrm{d}}
\newcommand{\eul}{\mathrm{e}}
\newcommand{\Cset}{\mathbb{C}}

\newcommand{\mt}{\mbold{g}}
\newcommand{\ma}{(\star)}
\newcommand{\maa}{(\star\star)}
\newcommand{\maaa}{(\star\!\star\!\star)}

\newcommand{\sot}{\mbox{$\,\,\mathop{\circledS}\,$}}

\newcommand{\ndx}[1]{\mbox{`$\mathstrut_{\!\hbox{\small\it #1}}$'}}

\newcommand{\dotabove}[1]{#1\kern-.5ex{\raise 1.7ex\hbox{.}}\,}
\newcommand{\Dotabove}[1]{#1\kern-.5ex{\raise 1.8ex\hbox{.}}\,}

\newcommand{\ndxx}[2]{\mbox{%
`$\mathstrut_{\!\hbox{\small\it #1}\dotabove{\hbox{\small\it #2}}}$'}}

\newcommand{\mTT}[2]{\vartheta_{\hbox{\small\it #1}%
                     \dotabove{\hbox{\small\it #2}}}}
\newcommand{\mbT}[2]{\vartheta_{#1\dot #2}}

\newcommand{\mbS}[2]{S_{#1#2}}
\newcommand{\mtS}[2]{S^{#1#2}}
\newcommand{\mdbS}[2]{S_{\dot #1\dot #2}}
\newcommand{\mSS}[1]{S_{\hbox{\small\it #1}}}

\newcommand{\tEps}[2]{{\epsilon^{#1 #2}}}
\newcommand{\dtEps}[2]{{\epsilon^{\dot #1\dot #2}}}
\newcommand{\bEps}[2]{{\epsilon_{#1 #2}}}
\newcommand{\dbEps}[2]{{\epsilon_{\dot #1\dot #2}}}

\newcommand{\mbG}[2]{\Gamma_{#1#2}}
\newcommand{\mbtG}[2]{\Gamma_{#1}^{#2}}
\newcommand{\mGG}[1]{\Gamma_{\hbox{\small\it #1}}}
\newcommand{\dmbtG}[2]{\Gamma_{\dot #1}^{\dot #2}}

\newcommand{\mbtO}[2]{\Omega_{#1}^{#2}}
\newcommand{\mbO}[2]{\Omega_{#1#2}}
\newcommand{\mbOO}[1]{\Omega_{\hbox{\small\it #1}}}
\newcommand{\mbW}[1]{\Psi_{#1}}
\newcommand{\mbWW}[1]{\Psi_{\hbox{\small\it #1}}}
\newcommand{\mbR}[4]{\Phi_{#1#2\dot #3\dot #4}}

\newcommand{\mbEM}[1]{\mbold{\phi}_{#1}}

\newcommand{\mtEM}[1]{\mbold{\phi}^{#1}}
\newcommand{\mdbEM}[2]{\mbold{\phi}_{\dot #1\dot #2}}

\newcommand{\mdtEM}[2]{\mbold{\phi}^{\dot #1\dot #2}}
\newcommand{\mbEEM}[1]{\mbold{\phi}_{\hbox{\small\it #1}}}
\newcommand{\mdbEEM}[1]{\mbold{\phi}_{\dotabove{\hbox{\small\it #1}}}}

\newcommand{\Vol}{V \kern -0.3em ol}

\newcommand{\EqN}[1]{\vb{Eq}.\ \vb{(#1)}} %

\newcommand{\npb}{\nopagebreak} 

\newcounter{line}

\newcommand{\fn}[1]{{\footnotesize\ref{line#1}}}

\makeatletter
\newcommand{\tlan}[4]{\addtocounter{line}{1}\relax%
                      \immediate%
                      \write%
                       \@auxout{%
                                \string\newlabel%
                                \string{line#2\string}%
                                \string{\string{\arabic{line}\string}%
                                \string{\arabic{page}\string}\string}%
                                }\relax%
                     \vspace{-#1em}\par\noindent\hphantom{xxxx}%
             \llap{\footnotesize%
                    #4{\it #3}%
                           }}
\makeatother

\newcommand{\tln}[2]{\addtocounter{line}{1}\relax%
                     \null\vspace{-#1em}%
                     \par\noindent\hphantom{xxxx}%
                     \llap{\footnotesize%
                           \hspace{-4pt}%
                           #2}}

\newcommand{\tlu}[1]{\null\vspace{-#1em}\par\noindent}

\newcommand{\tla}[2]{\tlan{#1}{#2}{input:\ }{\theline$\:$}}
\newcommand{\tlao}[2]{\tlan{#1}{#2}{output:\ }{\theline$\:$}}
\newcommand{\tlo}[1]{\tln{#1}{\it output:\ }}
\newcommand{\tli}[1]{\tln{#1}{\it input:\ }}
\newcommand{\tliz}[1]{\tln{#1}%
           {\mbox{$\cdot\!\cdot\!\cdot\,$}\it input: }}
\newcommand{\tloz}[1]{\tln{#1}%
           {\mbox{$\cdot\!\cdot\!\cdot\,$}\it output: }}
\newcommand{\tl}[1]{\tln{#1}{}}

\newcommand{\rbrk}{$\triangleleft$} 
\newcommand{\lbrk}{$\triangleright$} 

\newcommand{\rzr}{\hfill\strut\hfill%
                  $\wr\!\!\leftrightsquigarrow\!\!\wr$\hfill}

\newcommand{\filine}[1]{\tla{\firstgap}{#1}}
\newcommand{\siline}[1]{\tla{\gaps}{#1}}
\newcommand{\szline}{\tliz{\gaps}}
\newcommand{\foline}[1]{\tlao{\firstgap}{#1}}
\newcommand{\soline}[1]{\tlao{\gaps}{#1}}
\newcommand{\sozline}{\tloz{\gaps}}
\newcommand{\dfline}{\tl{\firstgap}}

\newcommand{\dfoline}{\tlo{\firstgap}}
\newcommand{\dsoline}{\tlo{\gaps}}
\newcommand{\dfiline}{\tli{\firstgap}}
\newcommand{\dsiline}{\tli{\gaps}}

\newcommand{\sline}{\tlu{\gaps}}
\newcommand{\bline}{\nopagebreak\tlu{\brgaps}\hfill\lbrk}

\newcommand{\droppingout}[2]{%
\null\vspace{-#1 em}\relax%
\par\noindent\hfil\hbox{\small$\lll$}\hfil
{#2}
\hfil\hbox{\small$\ggg$}\hfil\hphantom{.}\linebreak}

\newcommand{\dropout}[1]{\droppingout{\gaps}{%
\em the continuation dropped out consists of\/ {\rm #1} non-empty
lines
}}

\newcommand{\ovrl}{\hbox{\tt\char"7E}}

\newcommand{\und}{{\mbox{\tt\char'137}}}
\newcommand{\vrt}{{\hbox{\hskip -0.3ex\tt\char"7C\hskip -0.3ex}}}

\newcommand{\nt}[1]{{\mbox{$\langle{}#1{}\rangle$}}}
\newcommand{\ntt}[2]{\nt{{}#1{}$\und${}#2{}}}

\newcommand{\Rmr}[1]{{(\small{\em Remark:} #1)}}

\begin{document}

\title{\Large
Searching for electrovac solutions \\
to Einstein--Maxwell equations \\
with the help of computer algebra system \\ \GRG
}

\author{S.I.\ Tertychniy%
\thanks{bpt97\char"40{}ftri.extech.msk.su}\\
\normalsize
National Institute of Physical--Technical\\
\normalsize
and Radiotechnical Measurements\\
\normalsize
(VNIIFTRI),\\
\normalsize
Mendeleevo, Moscow region,
141570, Russia}

\date{}
\maketitle

\begin{abstract}
An example of application of the specialized computer
algebra system \Grg\ to the searching for solutions to the
source-free Maxwell and Einstein--Maxwell equations is
demonstrated. The solution involving five arbitrary functions of
two variables is presented in explicit form (up to
quadratures).

\end{abstract}

\section{Introduction}

 The specialized computer algebra system \Grg\ \cite{Grg}\ is
intended
for applications to the classical theory of
gravity as well as adjacent problems related to the classical theory of field
and geometry.
 In particular, \Grg\ is currently `aware' of the majority of main
characteristics of the geometry of a curved space utilized in
Einstein' theory. These are, for example, the bases in foliations of
exterior forms connected with metric, the connection (including its
Newman-Penrose representation), the curvature, its irreducible
constituents and algebraic invariants, the general equations
connecting these objects (Cartan' structural equations, Bianchi'
equations, various algebraic identities), the field equations of the
gravity theory (Einstein equations for vacuum and various matter
content). This list can be continued with the basic elements of the
Rainich theory, the theory of Lanczos potential, the methods of the
symmetry description (Killing equations). \Grg\ system operates with
major characteristics of such classical fields as electromagnetic
field, massless spinor field (Weyl field), massive spinor fields
(Dirac fields) including the both cases of the inclusion of interaction with
electromagnetic field (``charged" Dirac field) and the absence of
electromagnetic interaction, massless scalar field, conformally
invariant scalar field, massive scalar field, massive vector field
(Proca field), pressure-free dust matter, massive and null, \etc%
  \footnote{%
This list should not regarded as ultimate one.
}

 Concerning the above characteristic of \Grg\ capabilities, it has
to be noted that there is, of course, a number of programming
packages of symbolic manipulations with a similar
application field (an excellent survey of the latter topic is given
in Ref.~\cite{MacC}, see also \cite{Hart}). At the same time a
general overview of the current state of this field of computer
applications led a majority of authors to the conclusion that, in
spite of noteworthy achievements of the applied computer algebra, it
would be still premature now to single out any of the existing
programming systems of the class implied. None of them can be
assessed as an universal tool which reveals unconditional advantages
over all the potential competitors and best suits for majority of
applications. Accordingly, in practice, answering what is the best
system of symbolic calculations for application to the gravity
theory, one has to preliminary make definite what a class of
problems to be tackled is implied. This observation allows one to
suppose that \Grg\ can find an own position in the row of
computer-based tools utilized for theoretical investigations in the
field of the gravity theory.

It is worth mentioning that
 one may find a lot of examples giving evidence that the practical
usefulness of programming instruments intended to `non-local'
applications is a fairly subtle matter affected by a plenty
of factors, often rather vague and sometimes fairly unexpected. In
particular, it may be stated that any partial characteristic
(similar to the above list the physical-mathematical material
implemented in \Grg) or any verbal description is not able
to ultimately establish a proper measure of the actual advantages
--- as well as drawbacks --- of a software given.

 Nevertheless, there is an evident straightforward mean to gain some
insight into the heart of the problem. Specifically, an overview of
a proper set of manifest examples of the practical use of a
programming system can serve a probe exhibiting its real
characteristics. Concerning \Grg\ system, the present work realizes
a step just in that direction.

 In order to demonstrate in acting the basic features of \Grg\ system
(and to exhibit some of its capabilities) we would like to learn the
solving of a concrete problem, namely, the integrating of the
system of Maxwell and Einstein--Maxwell equations. The result we shall derive
is of a certain independent interest. The solution of the
electrovac field equations obtained below with the help of --- or, if
one prefer, in collaboration with --- \Grg\ system involves five
arbitrary functions of two variables. Though the calculation we
picked does not pretend to be related, with regard to its
complexity, to the category of record ones, it nevertheless gives, in
our opinion, a clear evidence of a usefulness of the programming tool
presented for the handling of such sort problems.

 It is worth noting that our choice of the subject of consideration
was severely conditioned by fairly simple `pragmatic' reasons.
On the one hand
it is evident that consideration of a toy problem would
obviously decrease a potential interest of the demonstration. In any
case, it tells a reader nothing definite in connection with the real
efficiency of the computer tool applied. On the other hand, in the
opposite extreme case of complicated problems
lying on the edge of the practical capabilities of the method, there
would appear an evident technical obstacle strongly restricting the
scope of discussion. Indeed, solutions of complicated problems
are usually too bulky to be exhibited in sufficient details within
frames of an article of any reasonable length%
 \footnote{It may not be definitely stated that, after all,
           this characteristic of the present paper proves to be
           unconditionally satisfactory
           but there is seen no ways to amend it in that respect.}.
In the best case the
only initial posing of the problem and the final result with
minimal discussion can be presented, the explanations of the essence
and, all the more, the features of the course of the solving
procedure being inevitably omitted. (As an indirect evidence
of such a state of things we
might refer to a lot of publications of investigations resorting to
assistance of the computer algebra where a remark "\dots with the help
of computer algebra we found that\dots'', or the like, is the only
`discussion' of the relevant programming issues.)

In our case,
 as usually, a contradiction of the choice among two opposite
extremes is to be settled by means of a compromise. Specifically,
although the problem we shall discuss below is in principle tractable by
means of a `manual' calculation (being rather error prone though),
definitely, nobody would qualify it at whole as a trivial one. It seems
thus to be able to provide a proper material for the displaying the
fashion of the work with the computer tool considered.

 Reverting to the general characterizing of \Grg\ system, it has to
be mentioned that one of the major goals pursued during its
development was an attempt to release a user from the duty to
`develop a program' --- at least, in the meaning usually associated
with the latter pursuit. Rather, dealing with \Grg\ system,
one has to merely describe, employing
mostly `habitual' words arranged in a sequence of natural
`phrases', and operating with more or less standard
mathematical notations, what initial data are given and indicate
what a result has to be generated. To that end, \Grg\ maintains
the input language which is, probably, as close as possible to the
one used for the representation of the relevant notions and the
relationships taking place in the application field itself. Of
course, comparatively sophisticated mathematical manipulations
require from a user an additional control over the calculations. In
particular, only he or she is able to decide what a way should be
most fruitful under the specific conditions depending on the details
of the method of the problem treatment and the data given. However
this, anyway, means that a researcher may focus on the essence of
the problem considered, the technical programming-related issues
requiring considerably less outlay than the other symbolic computational
tools would necessitate.

Currently, \Grg\ system enables one to calculate or to subject to
the other processing more than two hundreds of the so called {\em data
objects\/} modeling the basic notions and relationships (equations)
originated from the field theory in the curved space-time and the
geometry. However \Grg\ is {\em not\/} accommodated for
the purpose of the abstract index manipulating.
The handling of
data objects endowed with extrinsic indices (tensors, spinors, \etc)
is carried out utilizing, essentially, the explicit sets of their
components attributed to a definite gauge.

Touching in brief
the topic of realization, it is to be noted that
\Grg\ is a `superstructure' over the well known
general purposes computer algebra system \Red\ developed by A.~Hearn
(see Ref.~\cite{Hearn}). Accordingly, \Grg\ potentially possesses the
same degree of portability as \Red\ does.
The programming language utilized for the \Grg\
realization is not {\sl Rlisp}, the language of the own \Red\
coding, but the Lisp dialect named {\sc standard lisp}, see
Ref.~\cite{Psl}, which is supported in frames of the {\sl PSL}
({\sl Portable Standard Lisp}) package, the bottom level of the \Red\
infrastructure hierarchy%
 \footnote{It is worth noting that {\sl PSL} is not the exclusive
possible base suitable for the \Red\ implementation.}%
.
 It is important to emphasize that a user need be familiar with
neither {\sc standard lisp}, nor with \Red\ (although a knowledge of
the latter system would not be completely useless
of course especially in the
cases of more sophisticated calculations). Specifically, \Grg\
creates an `opaque casing' which mostly conceals the lower level
data structures utilized and the inner ways of their handling. \Red\ is thus
a `lower level' structure from viewpoint of \Grg.

 A language exploited for the communication of a user with \Grg\
system is a product of the independent working up. It is closely
related to the application field and, at the same time, mimics the
elements of the natural language, utilizing a sort of partial
simulation of English. Accordingly, the essence of \Grg\ programs
is, as a rule, understandable for any specialist in the application
field --- even a dilettante in the programming as such. In the worst
case, a quite moderate number of elucidations of specific issues
could prove to be necessary (one will see this on examples
considered in the present work).

The source code of \Grg\ amounts now to approximately 1.5 Mb. The
compiled (32--bit) binary code occupies about 1 Mb of a
disk space.
\Grg\ is currently intended for a free distributing%
.

\section{Preliminaries}\label{prel}

 It has to be noted that in the case of four dimensional Lorentz
geometry \Grg\ uses for calculations a version of the null tetrad
method based on Cartan' structural equations incorporated
with 2-spinors theory.

 Unfortunately, it is problematic to point out a single source
which would survey the formalism coinciding in all substantial
details with the one implemented in \Grg\ system%
 \footnote{This does not mean of course that the latter is unique in
           any respect.}%
. At the same time
the univocal representation of a mathematical matter underlaid
is of a notable
importance for anybody who deals with a concrete computer realization of a
mathematical formalism.

 Specifically, a well known detailed presentation (now, essentially,
unofficial standard) of the methods of spinor calculus is given
in the well known monograph
by Rindler \& Penrose \cite{RP}. However, at first, the
methods based on Cartan' structural equations and their
incorporating with spinor approach are not considered there in fact. At
second, apart from some notational discrepancies, there is a
distinction in the conventions fixing the sign of the spinor
contraction, \ie\ the definition what a rule, either
 $\iota_{A}=\bEps{A}{B}\,\iota^{B}$ or
 $\iota_{A}=\iota^{B}\,\bEps{B}{A}$,
 is to be used for spinor indices manipulations. The choices adopted in
\cite{RP} and in \Grg\ are opposite. [The point is that the both
above definitions were (and are, see, \eg, \cite[Appendix A:A]{CRPU})
utilized in the literature. The
former one is adopted, in particular, in the series of investigations
by J.~Pleba\`nski {\em et al\/} considering complex Einstein spaces,
\cf, \eg, \cite{PP}, Eq.\ (3.22).],
this system of notations and basic conventions
being elaborated a decade earlier than Ref.~\cite{RP},
adopting alternative choices, appears%
  \footnote{In connection with these remarks it should be mentioned that
            the predecessor of \grg\ is dated to 1981-2.}%
 .) Further, the methods of the calculus of exterior forms in a
version adapted to the gravity theory are outlined in Ref.~\cite{Cat}.
However, the spinor based description of the geometry
and field theory is given there after a fashion fairly incomplete
and insufficient for our purposes. We might also refer to the work
\cite{TO} but it treats a specific problem and cannot serve a review
of the formalism applied; besides, there are still some disagreements in
conventions utilized in Ref.~\cite{TO} and \Grg, unfortunately.

 We have to give therefore here a summary of the main
elements of the mathematical methods which will be `applied by \Grg\ system'
for the treatment of the problem considered in the main body of the
paper. Besides, we discuss here its mathematical posing necessary
for the subsequent processing by the methods of computer algebra.

\subsection{Elements of mathematical formalism}\label{math}

 Let 4-dimensional Lorentz metric be represented in the form of
expansion
\begin{equation}
\mt=\vartheta_{\phantom{A}\dot B}^{A\phantom{\dot B}}
\otimes
\vartheta_{A\phantom{\dot B}}^{\phantom{A}\dot B}.
                                                        \label{mdef}
\end{equation}
Here and below {\sl undotted spinor indices\/}  (denoted by upper-case Latin
{\small$ A,B$}, \etc) run over the two element set
$\{0,1\}$ while the {\sl dotted\/} ones (dotted upper case
{\small$ \dot A,\dot B$}, \etc) run
over the (distinct) set $\{\dot 0,\dot 1\}$ of `dotted integers'
(subjected, of course, the standard arithmetic). We use throughout the
standard spinor notations normalized in accordance with the following
convention which determines the spinor index manipulation rules \cite{PP}:
 $$
\iota_{A}=\bEps{A}{B}\,\iota^{B},\enskip
\iota^{B}=\iota_{A}\,\tEps{A}{B},\enskip
\iota_{\dot A}=\dbEps{A}{B}\,\iota^{\dot B},\enskip
\iota^{\dot B}=\iota_{\dot A}\,\dtEps{A}{B}{}.
 $$
Here 2-dimensional Levi--Civita symbols
 $\bEps{\star}{\star},\tEps{\star}{\star}$
may be defined as follows:
 \begin{eqnarray}
\bEps{A}{B}=-\bEps{B}{A},
\enskip\bEps{0}{1}=1;\quad
\tEps{A}{B}=-\tEps{B}{A},
\enskip\tEps{0}{1}=1;
                                              \nonumber\\
\dbEps{A}{B}=-\dbEps{B}{A},
\enskip\dbEps{0}{1}=1;\quad
\dtEps{A}{B}=-\dtEps{B}{A},
\enskip\dtEps{0}{1}=1.
                                              \label{eps}
 \end{eqnarray}
Further,
 $\vartheta_{A\dot B}$ (involved in Eq.\ (\ref{mdef}) with distinct
positions of indices) is the tetrad of the, generally speaking,
complex-valued covectors (1--forms) which are null with regard to
the metric $\mt$.
 In the case of the metric signature
$(-+++)$ assumed throughout
they always can be picked on to satisfy
the Hermitean-like symmetry  relation
 \begin{equation}
\mTT{0}{0}=\overline{\mTT{0}{0}},\enskip
\mTT{1}{1}=\overline{\mTT{1}{1}},\enskip
\mTT{0}{1}=\overline{\mTT{1}{0}}.
                                                \label{regauge}
 \end{equation}
 In such a case, the two tetrad elements $\mTT{0}{0},\mTT{1}{1}$ are real
while $\mTT{0}{1},\mTT{1}{0}$ are necessarily complex. (The meaning of
distinction in the format of numerical indexing which is observed in Eqs.\
(\ref{eps}) and (\ref{regauge}) will be explained below.)

We shall refer to Eq.\ (\ref{regauge}) as the {\em real gauge
condition}.
The imposing of this gauge does not completely fixes the
tetrad: it may be still undergone gauge transformations whose set
includes, in particular, the proper orthochronous Lorentz group.

Further,
it is sometimes convenient to utilize another indexing of the tetrad.
Let us consider the following re-definition%
 \footnote{Within framework of the matrix language
  this point assumes introduction of Pauli matrices.
  We do not utilize them in an explicit form.
}:
 \begin{equation}
\vartheta^{0}=\mTT{0}{1},\enskip
\vartheta^{1}=\mTT{1}{0},\enskip
\vartheta^{2}=\mTT{0}{0},\enskip
\vartheta^{3}=-\mTT{1}{1}.
                                                        \label{ts}
\end{equation}
In terms of such an indexing Eq.\ (\ref{mdef}) takes the form
\begin{equation}
\mt=2\,\vartheta^{0}\sot\vartheta^{1}+2\,\vartheta^{2}\sot\vartheta^{3},
                                                        \label{metr}
\end{equation}
where $\alpha\!\sot\!\beta
\equiv \half\alpha\otimes\beta+\half\beta\otimes\alpha$
denotes the symmetrizied tensor product.
We shall say that $\vartheta_{A\dot B}$ is the tetrad endowed with
{\em spinor indices\/} while Eq.\ (\ref{ts}) defines the {\em
tensorial\/} (or tensor) tetrad indexing.

Important algebraic objects which characterize the property of
space-time metricity are the two families of the so called
$S$-forms constituting 2-index spinors. In particular,
 {\it undotted $S$-forms\/} are defined as follows:
 \begin{equation}
 \mbS{A}{B}=\half\,\dtEps{K}{L}
 \mbT{A}{K}\wedge\mbT{B}{L}
 =\half\,\theta_{A\phantom{\dot{A}}}^{\hphantom{A}\dot{L}}
 \wedge\mbT{B}{L}^{\mathstrut}.
                                                        \label{Sform}
 \end{equation}

Next,
each metric induces a unique torsion-free
metric compatible connection. Within framework assumed, it is
described by the object
endowed with a pair of symmetric
undotted indices (which do not constitute a true spinor however)
whose components are 1-forms.
It is named undotted connection and is
denoted $\mbG{A}{B}$. Due to indicial symmetry there are
maximally {\em three\/} distinct components of undotted connection.
They are denoted
$\mGG{0},\mGG{1},\mGG{2}$.

 Notice that we used above so called {\em summed spinor indexing\/}
which applies to any object {\it symmetric\/} with respect to all
spinor indices of the same class (undotted and/or dotted ones)
including the case of connection. Hereinafter, dealing with numeric
values of spinor indices in explicit (expanded) formulae, the summed
indices are used throughout (see Eqs.\ (\ref{regauge}) and all the
subsequent ones; Eqs.\ (\ref{eps}) are an exception). In
particular, in the case of the undotted connection the complete
collection of its components comprises $\mbG{0}{0}\equiv\mGG{0},
\mbG{0}{1}=\mbG{1}{0}\equiv\mGG{0}, \mbG{1}{1}\equiv\mGG{2}$ (here
at right hand side the summed indexing is used while at left the
corresponding `numerical values' of `ordinary' spinor indices are
substituted, \cf\ Eqs. (\ref{eps}))%
 \footnote{As far as we know the simple but very useful notion of
summed spinor indices was never explicitly introduced and used on a
regular ground.}.

Given null tetrad,
the connection forms can be computed from the so called {\em first
structural equations\/} which are also
named the {\em first Cartan equations}.
There exist two their representations. The {\em first
version\/} of the first structural equations involves the
differentials of the tetrad elements and, in the case of the
fulfillment of the real gauge condition, reads
 \begin{equation}
\dex \theta_{A\dot B}^{\vphantom{\dot B}}
+\Gamma^{C}_{A\vphantom{\dot B}}
\wedge\theta_{C\dot B}^{\vphantom{\dot B}}
+\overline{\Gamma^{C}_{B\vphantom{\dot A}}}
\wedge\theta_{A\dot C}^{\vphantom{\dot B}}
=0.
                                                \label{feq1}
\end{equation}
Here the complex conjugation varies the class of connection indices,
\ie\ $\overline{\Gamma^{C}_{B\vphantom{\dot A}}}$ enters the
equation, effectively, with two {\em dotted\/} indices, {\small$ \dot
B$} and {\small$ \dot C$}, which are connected with undotted {\small$
B$} and {\small$C$}, respectively, by means of the map sending
 \hbox{\small\it 0} to \hbox{\small\it\Dotabove 0},
 \hbox{\small\it 1} to \hbox{\small\it\Dotabove 1} and \vv{}.
[More precisely, the so called {\em dotted connection forms\/}
$\dmbtG{B}{C}$, which are in general case algebraically independent on
undotted ones,
should be situated in Eq.\ (\ref{feq1}) instead of
$\overline{\mbtG{B}{C}}$.
However, under the real gauge condition, $\dmbtG{B}{C}$ and
$\mbtG{B}{C}$ are mutually `Hermitean conjugated' and we
prefer to evade introduction of an additional object which is, essentially,
superfluous for our current purposes.]

The {\em second version\/} of the first structural equations
(first Cartan equations) involves
differentials of $S$-forms instead of ones of tetrad
elements and reads:
 \begin{equation}
\dex\mbS{A}{B}^{\vphantom C}
+2\,\Gamma^{C}_{(A}\wedge S_{B)C}^{\hphantom{C}}=0.
                                                \label{feq2}
 \end{equation}
The both versions of the first structural equations (which are,
essentially, equivalent) always can be algebraically resolved with
respect to the connection forms.

The next important geometric relationship utilizes the so called {\em
undotted curvature spinor\/} $\mbO{A}{B}=\mbO{B}{A}$ whose each
component is a 2-form. The link of undotted
curvature and undotted connection is yielded by the {\em second
structural equation\/}%
  \footnote{undotted one, there is also dotted second
            structural equation
            involving dotted curvature and dotted connection}.
It reads
 \begin{equation}
\mbtO{A}{B}=\dex\,\mbtG{A}{B}+\mbtG{A}{K}\wedge\mbtG{K}{B}.
                                                \label{seq2}
 \end{equation}
Space-time is {\em locally flat\/} if and only if $\mbtO{A}{B}$
(and its dotted counterpart) vanishes.

The specific relationships
lying in foundation of the general relativity
(Einstein equations for the general matter content and their particular
case, Einstein--Maxwell equations, which we shall deal with in the
present work alone) are formulated making use of the so called {\em irreducible
spinor curvatures\/}.
The latter involve in particular, the {\em undotted Weyl spinor\/}
$\mbW{ABCD}=\mbW{(ABCD)}$ representing
Weyl curvature tensor and its Hodge dual, {\em Ricci spinor\/}
$\mbR{A}{B}{C}{D}=
\mbR{(A}{B)\,}{{{\raisebox{0pt}[0pt][0pt]{\makebox[0pt]{\scriptsize(}}C}}}{D)}$
equivalent to the traceless part of
the Ricci curvature tensor, and the {\em scalar curvature\/} $R$. All these
objects can be determined from the known undotted curvature (and
$S$-forms) with the help of the equations displayed below:
 \begin{eqnarray}
&&\imu\,\mbW{ABCD}\cdot\Vol
=
\mbO{(A}{B}\wedge\mbS{C}{D)},
                                             \quad
-\imu\,\mbR{A}{B}{C}{D}\cdot\Vol
=
\mbO{A}{B\vphantom{\dot D}}\wedge\mdbS{C}{D},
                                             \label{spc}
                                                      \\
&&
\imu\,R\cdot\Vol=4\mbO{A}{B}\wedge\mtS{A}{B}.
                                             \nonumber
\end{eqnarray}
Here the dotted $S$-forms $\mdbS{C}{D}$ are defined similarly to
undotted ones by the equation
$$
\mdbS{A}{B}=\half\,\tEps{K}{L}
\mbT{K}{A}\wedge\mbT{L}{B}
$$
(\cf\ Eq.\ (\ref{Sform}))
and
$$
\Vol=\Ratio{\imu}{12}
\theta_{A\vphantom{\dot B}}^{\dot B}\wedge
\theta_{\dot B}^{C}\wedge
\theta_{C\vphantom{\dot B}}^{\dot D}\wedge
\theta_{\dot D}^{A}$$
is the volume element (nonzero 4-form) induced by the corresponding
metric.

 Now let us outline the form of the basic equations of
electromagnetic theory within framework of the formalism under
consideration. They are extremely transparent.
Electromagnetic field is associated with the
complex-valued 2-form $\omega$ which is named {\em undotted
electromagnetic 2-form}. It has to be spanned by the undotted
$S$-forms, \ie{}, admits the expansion
 \begin{equation}
\omega=\mtEM{AB}\mbS{A}{B}.
                                         \label{asd}
 \end{equation}
The above coefficients $\mbEM{AB}=\mbEM{BA}$ constitute the so called
{\em undotted spinor of electromagnetic field}.
The charge- and current-free Maxwell equations reduce to the
closeness condition imposed on the undotted electromagnetic 2-form:
 \begin{equation}
  \dex\,\omega=0.
                                                \label{max}
 \end{equation}
We shall refer to it as {\em undotted Maxwell equations}. In the case
of the imposing of the real gauge condition (\ref{regauge}) the
complex conjugation of the undotted electromagnetic 2-form
yields the {\em dotted electromagnetic 2-form\/}
 $\dot\omega=\bar{\omega}$ which admits a similar expansion
 $$
\dot\omega=\mdtEM{A}{B}\mdbS{A}{B},
 $$
where the coefficients $\mdbEM{A}{B}$ constitute the
{\em dotted spinor of electromagnetic field\/}.
Finally, under a suitable choice of the physical
units, Einstein--Maxwell equations are represented
as follows:
 \begin{eqnarray}
2\mbR{A}{B}{C}{D}&=&\mbEM{AB\vphantom{\dot C}}\mdbEM{C}{D},
                                                        \label{spp}\\
R&=&0.
                                                        \label{scp}
\end{eqnarray}
Eq.\ (\ref{spp}) is called the {\em spinor part of
Einstein--Maxwell\/} equations while (\ref{scp}) is their {\em scalar
part\/}.

Now we may proceed with the description of the problem
chosen for the demonstration of the application of the
formalism outlined with the help of \Grg.

\subsection{Radiative electrovacs}\label{s2}

In the present work, we shall consider solutions of the
source-free Maxwell and Einstein--Maxwell equations ({\em
electrovacs}) which, under a proper gauge, satisfy the equation
 \begin{equation}
  \mGG{0}=0.
                                                     \label{raev}
 \end{equation}
Characterizing its geometric,
the first important consequence follows from
\ndxx{0}{0}--component of Eq.\ (\ref{feq1}) which now reads
\begin{equation}
\dex\,\theta^{2} +2\,\theta^{2}\wedge\re\mGG{1}=0.
                                                 \label{feq}
\end{equation}
Specifically, by virtue of Frobenius' theorem, $\theta^{2}$ determines
an integrable distribution. Said another way, there exists, at least locally,
a scalar function $x$ such that
\begin{equation}
\theta^{2}=\Xi_1\,\dex\,x
                                                 \label{xeq}
\end{equation}
for some non-zero factor $\Xi_1$, the hypersurfaces $x=$constant being null.

Further, it has to be noted that the coefficients
$\rho,\sigma,\dots\pi$
of the general expansion of undotted connection with respect to null tetrad
 \begin{eqnarray}
\mGG{0}&=&\rho \theta^{0}+\sigma \theta^{1}
        +\tau \theta^{2} - k\theta^{3},
                                       \nonumber 
                                                      \\
\mGG{1}&=&\alpha \theta^{0}+\beta \theta^{1}
        -\gamma \theta^{2} + \varepsilon \theta^{3},
                                       \label{eq03200}
                                                      \\
\mGG{2}&=&\lambda \theta^{0}+\mu \theta^{1}
        -\nu \theta^{2} + \pi\theta^{3}.
                                       \nonumber 
\end{eqnarray}
 are nothing else but the Newman-Penrose (NP) spin coefficients, see
Ref.~\cite{NP}, \cite{Cat}. Some of them (namely those involved in
expansion of $\mGG{0}$ and $\mGG{1}$) describe the invariant
properties of the null congruence determined by the covector
$\theta^{2}$, \ie, generating hypersurfaces $x$=constant. In
particular, if $k$=0 (that follows from (\ref{raev}))
this congruence is geodetic and possesses the
complex expansion equal to $\rho$ and shear $\vert\sigma\vert$. Thus
we immediately
see that all the optical scalars vanish in the case under
consideration.

Of course
the ansatz (\ref{raev}) imposes severe restrictions on the
space-time curvature as well. Specifically, the
\ndx{0}--component of Eq.\ (\ref{seq2}) reads
 \begin{equation}
\mbOO{0}=\dex \mGG{0}+2\mGG{0}\wedge\mGG{1}
                                          \label{Om0}
 \end{equation}
and, thus, Eq.\ (\ref{raev}) immediately yields $\mbOO{0}=0$. This,
in turn, implies in particular the following constraints on the
components of the Weyl spinor and the scalar curvature
(see Eq.\ (\ref{spc})):
 \begin{equation}
0=\mbWW{0}=\mbWW{1}=\mbWW{2}+\Ratio{1}{12}R.
                                            \label{III}
 \end{equation}
Since in electrovac space-times the scalar curvature $R$ vanishes
(Eq.\ (\ref{scp})), $\mbWW{2}$ vanishes as well
and only $\mbWW{3}$, $\mbWW{4}$ may be non-zero.

 The above restriction on the components of Weyl spinor implies that
the space-times in question belong to Petrov--Penrose types {\sf
III} or {\sf N} (or, perhaps, are conformally flat, the type {\sf
O}). The corresponding multiple principle null direction coincides
with the one determined by the congruence of null geodesics
generated by null hypersurfaces $x=$constant. We have seen that all its
optical scalars vanish.

 It is known that
gravitational fields with the above
algebraic structure of the conformal curvature are usually associated with
gravitational radiation spread in the 4--dimensional light-like
direction, multiple principal direction. In particular, the
parallelly propagated plane waves ({\em pp}--waves, see Ref.~\cite{Cat},
section 21.5) satisfy all the above conditions. Due to
these reasons and for convenience of further references, we shall
name the solutions of the source-free Maxwell and Einstein--Maxwell
equations satisfying the restriction (\ref{raev}) {\em radiative
electrovacs}%
 \footnote{It should be mentioned however that there exist
   configurations describing radiation which do not belong to the class
   singled out. A wellknown example is, in particular, Robinson-Trautman'
    class of metrics, see Ref.~\cite{Cat}.
On the other hand, there are `radiative' spacetimes (in our terminology)
which are unlikely to be associated with any radiation process,
\eg, the Hauser' solution \cite{Hau}.
}.

 In the present work, we find the `most general' metric belonging to
the class of radiative electrovacs, reducing all the field equations
to several subsequent quadratures. Our solution does not actually
includes all the radiative electrovacs but the metrics missed here
(\eg\ {\em pp}--waves) are either generated from it by means of
appropriate limiting processes or arise following a particular
`degenerated' side-branches of the `generic' integration procedure.

Now we are able to carry out a complete posing of the problem.

\subsection{Posing of the problem}\label{s3}

 First of all we have to determine what a tetrad fits the
restriction (\ref{raev}). To that end, we shall utilize
some further implications
inferred by Eq.\ (\ref{feq1}). Specifically, it has been mentioned
that its \ndxx{0}{0}--component implies Eq.\ (\ref{xeq}). Absorbing
the factor $\Xi_1$ by the tetrad element $\theta^{3}$ (which is an
admissible gauge transformation, \cf\ Eq.\ (\ref{metr})), we obtain
that, without loss of generality, one may assume
\begin{equation}
\theta^{2}=\dex\,x/\ssqrt{2}.
                                                \label{th2}
\end{equation}

 Further, eliciting
the \ndx{0}--component of Eq.\ (\ref{feq2}),
one obtains
$$ 
\dex\,\mSS{0}   +2\,\mSS{0}\wedge\mGG{1}=0,
$$ 
where $\mSS{0}=\theta^2\wedge\theta^0$
(see Eqs.\ (\ref{ts}), (\ref{Sform})).
Again, the Frobenius' theorem implies
$$\mSS{0}=\eul^{\alpha}\dex\,\hat x\wedge\dex\,\phi$$
 for some functions $\alpha,\hat x,\phi$. By virtue of (\ref{th2}),
without loss of generality, one may get $\hat x=2\,x$
(we \apriori\ know that by virtue of definition $\mSS{0}=\theta^2\wedge\dots=
        (\dots)\dex\,x\wedge\dots$)
and, then,
executing, if necessary, some gauge transformation, we obtain
\begin{equation}
\theta^{0}=\eul^{\alpha}\dex\,\phi/\ssqrt{2}\enskip
\Rightarrow\enskip
\theta^{1}=\overline{\theta^{0}}=
           \eul^{\bar\alpha}\dex\,\bar\phi/\ssqrt{2}.
                                                \label{th01}
\end{equation}
 It is immediately clear from Eq.\ (\ref{metr}) that the only value
of $(\re\alpha)$ is significant while the choice of $(\im\alpha)$ is
a matter of gauge. Indeed $(\im\alpha)$
may be arbitrary, leaving the metric
unaffected. Additionally, analyzing the above explicit equations, it
is easy to see that the local coordinate transformation
$\phi\rightarrow\Xi_2(\phi,x)$ for arbitrary $\Xi_2$ holomorphic
with respect to $\phi$ (and satisfying the natural restriction
$\Xi_2\strut,_\phi\ne0$%
 \footnote{Hereinafter, a subscript
          following comma denotes partial derivative.}%
 ) still remains allowable, provided the corresponding
transformation of $\alpha$ accompanies it. We reserve this
possibility for a future use.

Noting that
$\theta^0\wedge \theta^1\wedge \theta^2\ne0$, the variables
$x,\phi,\bar\phi$ are functionally independent (and thus $\phi$ is
complex and `functionally independent' of $\bar\phi$,
\ie\ $\phi$ and $\bar\phi$ are a form of representation
of two real independent coordinates).
Let us use them as three of four coordinates and denote the
fourth independent real coordinate $y$.

Thus, up to now, the only non-specified tetrad element is
$\theta^3$. Its generic holonomic expansion reads
\begin{eqnarray}
 \ssqrt{2}\:\theta^3&=&\Xi_3\,\dex y+2\re(\chi\,\dex\,\phi) + h\,\dex\,x,
                                 \nonumber
\end{eqnarray}
where the coefficient $\Xi_3$ may not vanish,
$\chi$ (complex) and $h$ (real) are some functions.
 Then, executing the admissible change of the coordinate $y$,
$y\rightarrow \int \Xi_3\,\dex y$, one always may reduce (the
counterpart of) $\Xi_3$ in new coordinates to a constant. Thus we
may assume
 \begin{equation}
 \ssqrt{2}\:\theta^3=\dex y+2\re(\chi\,\dex\,\phi) + h\,\dex\,x.
                                 \label{th3}
\end{equation}
Let us notice further that the transformation
\begin{equation}
 y \rightarrow y+\Xi_4
\end{equation}
for arbitrary real function
$\Xi_4=\Xi_4(x,\phi,\bar\phi)$
which is accompanied with the replacing
\begin{equation}
\chi \rightarrow \chi - \Xi_4\strut,_\phi, \qquad
\bar\chi \rightarrow \bar\chi - \Xi_4\strut,_{\bar\phi}
\end{equation}
 and some transformation of $h$ is also a change of a gauge. (It
does not affect any of equations considered above including the expansion
(\ref{th3}).) Notice also that then we have the following subsidiary
transformation;
 \begin{equation}
\chi\strut,_{\bar\phi}
+
\bar\chi\strut,_\phi
\rightarrow
(\chi\strut,_{\bar\phi}
+
\bar\chi\strut,_\phi)
- 2\Xi_4\strut,_{\phi\bar\phi}.
                                        \label{chigau}
\end{equation}
It will be useful below.

 Ultimately, we can formulate the problem as follows: We shall
search for electrovacs whose metric is described by the null tetrad
(\ref{th01}), (\ref{th3}) and possesses connection which satisfies
the constraint (\ref{raev}).

\section{Handling field equations with the help of computer
algebra}\label{grgappl}

 In this section subdivided into several subsections we present the
application of \Grg\ system for the reduction of the field equations
specifying radiative electrovacs. In each step, a separate series of
calculations is carried out yielding some new information describing
the structure of solution. Whenever its `amount' exceeds some
reasonable threshold, the calculation stops and further proceeds with
the next step from the beginning, utilizing the relationships
obtained so far as a new and more detailed initial data.

Any knowledge of \Grg\ system is not presupposed.
Accordingly, relevant explanations on this issue are given whenever
necessary%
 \footnote{excluding some general means of the formulae coding which
are utilized in majority of programming languages such as Fortran,
Pascal, \etc},
making the discussion, essentially, self-contained.

\subsection{Step 1: Coding of the initial data and inference of
first implications
}\label{start}

Generally speaking,
 the data to be read and processed by \Grg\ is to be arranged in a form of
the so called {\em problem specification\/} which is a regular text
usually stored in a disk file.
(One finds in Appendix A a sample of the workable problem
 specification.)
The structure of problem specification
is fairly transparent. Apart from a title and conclusion, it
consists of a series of {\em sections\/} comprised, in turn, of
several {\em paragraphs\/}
and must, in particular,
contain the section of (initial) \vb{DATA}. The other sections
(for example, the section of \vb{INSTRUCTIONS}, the section of
\vb{SUBSTITUTIONS}, and some others) are optional.

 Specifying initial \vb{DATA}, a user has to explicitly list all
the non-implemented (`user-defined')
objects which he or she is going to utilize.
 Here the particular symbols of \vb{COORDINATES}, specific
\vb{FUNCTIONS} (`unknown' to the system),
\etc, are, in particular, meant%
  \footnote{One may organize a library of such sort data
and load them from there.
}%
.
In our case all what we need at the beginning is explicitly
\vb{DECLARE}d and defined as follows:
\filine{1}\verb?Data:?
\dsiline\verb? declare COORDINATES x,y,phi,phi~;?
\siline{3}\verb? declare SCALARS al,chi,h(x,y,phi,phi~);?
\dsiline\verb? declare REAL x,y,h;?
\siline{5}\verb? declare COMPLEX CONJUGATED phi & phi~,al & al~,chi & chi~;?
\someskip
\siline{6}\verb? TETRAD comprises?
\szline\verb?    component|0=E**al d phi/sqrt(2),?
\siline{8}\verb?    component|1=C.C.(component|0),?
\szline\verb?    component|2=d x/sqrt(2),?
\siline{10}\verb?    component|3=(d y +2 RE(chi d phi)?%
\verb? +h d x)/sqrt(2);?
\someskip
\siline{11}\verb?end of data.?
\par\noindent
 Here the symbol \Vb{al} stands for $\alpha$ (see Eq.\
(\ref{th01})), \Vb{C.C.} is the symbol of the {\em complex
conjugating\/} operator while the origin of the other symbols seems
to be evident. Continuing with a matter of notations, it is also useful
to mention that here and below the enumeration at left hand side in the copies
of fragments of \Grg\ input
scripts and output listings, as well as
the marks at left `{\small\it input:}' (line of input code),
 `{\small\it output:}'
(line of output file), \etc, are not a part of the texts displayed. They
were introduced for convenience of further references alone. For example,
we can refer to the lines
\fn{1},\fn{3},\fn{5},\fn{6}--\fn{10},\fn{11} involved in the
\vb{DATA} section displayed above.)

 It is seen that the above script encodes, almost
verbatim, the content of Eqs.\ (\ref{th2}, \ref{th01}, \ref{th3}).
Nevertheless a number of comments, concerning some specific points of
\Grg\ input language, should be made.

First, the tilde symbol `\verb?~?' attached at right to some
identifiers imitates the overscoring used to denote complex
conjugation in the standard system of mathematical notations. For example,
\Vb{phi\ovrl} may be regarded as the representation of $\bar\phi$,
 \etc{}
 However, contrary to the habit of mathematical notations, the
additional tilde mark does not impose itself any actual relation
between the corresponding identifiers. For example, \vb{phi} and
\vb{phi}\ovrl\ are \apriori\ not connected in any way. The encoding
of the necessary relationship such as $C.C.(\phi)=\bar\phi$ is realized by
means of a separate declaration. See line {\fn{5}}.

Next, the term \Vb{SCALARS}%
    \footnote{It has no relation to the own \Red' \vb{scalar}s.}
denotes, roughly, a sort of functions each
of whose is at the same time a single symbol
(identifier). The dependence upon the (fixed) set of arguments%
 \footnote{which, to be more precise, is varied by the coordinate
           transformations, if any}
pointed out in the declaration (line \fn{3}) is
supported by a special implicit mechanism.

 We also mention that the symbol \vb{df} (or, equivalently, \vb{DF})
denotes partial derivative (as well as ordinary derivative of a
function of a single argument which is not formally distinguished
from a partial derivative). In particular, \vb{df(chi,phi)} encodes
$\partial\,\chi/\partial\,\phi$. The derivative
$\partial^2\,\chi/\partial\,\phi/\partial\,\bar\phi$ is represented
by the script \vb{df(chi,phi,phi\ovrl)}, for
$\partial^2\,\chi/\partial\,\phi^2$ one has to write down
\vb{df(chi,phi,2)},
\etc\hphantom{.}
(These notations are identical to ones supported by \Red\ system.)

Finally,
 the record `\vb{component}\vrt\nt{j}' (the delimiter \vb{\vrt} is
here optional and might be dropped out), where \nt{j} is a digit,
refers to a single component of a data object, in our case
\vb{TETRAD}. Thus the record
 `\vb{component}\vrt\vb{1}\vb{=}
\vb{C.C.(component}\vrt\vb{0)}' (see line \fn{8}) represents the equation
$\theta^0=\overline{\theta^1}$, \cf\ Eqs.\ (\ref{regauge}),
(\ref{ts}).

Now we are ready to proceed with calculations.

 Let \Grg\ package was started and has read the problem
specification consisting of a single section of \vb{DATA}. We may to
immediately control the calculations issuing the {\em
instructions\/} from a keyboard. (Alternatively, they --- or a part
of them --- might be put into a section of \vb{INSTRUCTIONS},
provided it were included in the problem specification. The mixed
form of control is also available.)

Thus, at first, after the auxiliary request to
\dfiline\verb?turn on displaying of negative powers?%
        \footnote{which may be shortened to \Vb{turn on DIV}}
\par\noindent
affecting the format of output,
we issue the instruction to
\dfiline\verb?find UNDOTTED CONNECTION?
\par\noindent
The system response to it comes to the following output:
\dfoline\verb?==> find UNDOTTED CONNECTION?
\someskip
\dsoline\verb?UNDOTTED CONNECTION is not found in the DATA section...?
\someskip
\dsoline\verb?...UNDOTTED CONNECTION has been calculated?\rzr\verb?1.0 s.?
\someskip
\dsoline\verb?Total time spent amounts to 3.7 seconds.?
\par\noindent
\Rmr{
Hereinafter, the symbol
`{\normalsize\mbox{$\wr\!\!\leftrightsquigarrow\!\!\wr$}}'
denotes some blank space cut off from the copy of a line of
an input script or an output listing in order to
fit it to the page width.%
}

 However, we want to look at the result (\vb{UNDOTTED CONNECTION}) in
explicit form. To that end, we have to initiate the {\em action\/}
\Vb{TYPE} directed to the {\em data object\/} we are interested in.
The {\em shortened form\/} of the corresponding instruction
looks as
follows:
\dfiline\verb!UGAMMA0?!
\par\noindent
 Here \vb{UGAMMA0} is the reference to the component $\mGG{0}$ of
the \vb{UNDOTTED CONNECTION} (\Vb{UGAMMA} is its {\em kernel name\/}
while \Vb{0} is the value of the {\em summed undotted
index\/} distinguishing the component, see section \ref{math}).
 The return of the instruction
reads:
\dfoline\verb!==> UGAMMA0?!
\someskip
\dsoline\verb?--> TYPE UGAMMA0?
\someskip
\dsoline\verb?                  1    - al~?\npb
\dsoline\verb?  UGAMMA  =  ( - ---*E      *DF(chi~,y)) d x +  (?\npb
\dsoline\verb?        0         2?
\someskip
\dsoline\verb?                    1   al?\npb
\dsoline\verb?                - (---*E  )*(DF(al,y) + DF(al~,y))) d phi?\npb
\dsoline\verb?                    2?
\par\noindent
 Thus $\mGG{0}$ does not automatically vanish. This simply means
that the expression for the \vb{TETRAD} assumed is more
general than we actually need. At the same time the condition
ensuring the vanishing of $\mGG{0}$ is very simple: \vb{al+al\ovrl}
(\ie\ $\alpha+\bar\alpha$) and \vb{chi\ovrl} (\ie\ $\bar\chi$) have
to be independent on \vb{y} ($y$). Additionally, we have seen above
that the value of \vb{al-al\ovrl} is completely
gauge dependent. In
particular, it always may be assumed to be independent on \vb{y} and,
then, both \vb{al} and \vb{al\ovrl} turns out to be independent on
\vb{y} as well. Besides, $\chi$ is also independent on \vb{y}.
  We `inform' the system of these facts by means of the following instruction:
\dfiline\verb?Let DF(al,y)=0,DF(al~,y)=0,DF(chi,y)=0,DF(chi~,y)=0?
\par\noindent
The response reads
\dfoline\verb?==> Let DF(al, y) = 0, DF(al~, y) = 0, DF(chi, y) = 0,?%
\verb? DF(chi~, y) = 0?
\someskip
\dsoline\verb?   Section of SUBSTITUTIONS is not found.?
\dsoline\verb?   It is added to the problem specification.?
\someskip
\dsoline\verb? The following substitution rules?
\someskip
\dsoline\verb?        (1) DF(al, y) -> 0?
\dsoline\verb?        (2) DF(al~, y) -> 0?
\dsoline\verb?        (3) DF(chi, y) -> 0?
\dsoline\verb?        (4) DF(chi~, y) -> 0?
\someskip
\dsoline\verb?       are put to the section of SUBSTITUTIONS.?
\someskip
\dsoline\verb?  Substitutions 1,2,3,4 are now active?
\par\noindent
Now we can inspect the resulting structure of \vb{UNDOTTED CONNECTION}.
The instruction
\filine{1exf}\verb?evaluate UGAMMA0,UGAMMA1/\d x/\d phi/\d phi~,UGAMMA2/\d x/\d phi~?
\par\noindent
yields in particular the output
\dfoline\verb? The expression?
\dsoline\verb?      UGAMMA0?
\dsoline\verb?                  vanishes.?\rzr\verb?0.7 s.?
\someskip
\dsoline\verb?  The expression?
\soline{10g}\verb?      UGAMMA1/\ d x/\ d phi/\ d phi~?
\soline{11g}\verb?                  vanishes.?\rzr\verb?0.8 s.?
\someskip
\dsoline\verb?  The expression?
\soline{20g}\verb?      UGAMMA2/\ d x/\ d phi~?
\soline{21g}\verb?                  vanishes.?\rzr\verb?0.8 s.?
\someskip
\dsoline\verb? Total time spent amounts to 4.5 seconds.?
\par\noindent
 Thus the condition distinguishing radiative electrovacs (Eq.\
(\ref{raev})) is now satisfied. Additionally, we see that the holonomic
expansion of \vb{UGAMMA1} does not involve \vb{d y}-term while the
expansion of \vb{UGAMMA2} consists of the \vb{d x}- and \vb{d
phi\ovrl}-terms alone. We shall make use of these facts below.
 (We might display
the explicit expressions of all the components of \vb{UNDOTTED
CONNECTION} but these would be, essentially, useless here).

Our next operation implies the inferring of some information from the
field equations. Specifically, let us ask the system to
\filine{2k}\verb?obtain and isolate?
\szline\verb?   the SCALAR PART of EINSTEIN-MAXWELL EQUATIONS;?
\siline{4k}\verb?show EQUATIONS?
\par\noindent
The answer is the following:
\dfoline\verb?==> obtain and isolate the SCALAR PART of EINSTEIN ?%
\verb?- MAXWELL EQUATIONS?
\someskip
\dsoline\verb? --> OBTAIN SCALAR PART of EINSTEIN - MAXWELL EQUATIONS?
\someskip
\dsoline\verb? ...SCALAR PART OF EINSTEIN - MAXWELL EQUATIONS ?\rbrk
\bline\verb?has been obtained       4.0 s.?
\someskip
\dsoline\verb? Total time spent amounts to 8.6 seconds.?
\someskip
\dsoline\verb? --> ISOLATE SCALAR PART of EINSTEIN - MAXWELL EQUATIONS?
\someskip
\dsoline\verb?                           al  al~?\npb
\dsoline\verb? Warning: The denominator E  *E    has been removed?%
\verb? assuming it to be?
\someskip
\dsoline\verb?           nonzero...?
\someskip
\dsoline\verb?    EQ. number 1 is put to the list of EQUATIONS.?
\someskip
\dsoline\verb? Total time spent amounts to 8.7 seconds.?
\someskip
\dsoline\verb? ==> show EQUATIONS?
\someskip
\dsoline\verb?                      al + al~?\npb
\dsoline\verb?  Eq. (1) reads 0=4*(E        *DF(h,y,2) - DF(al,phi,phi~)?
\someskip
\dsoline\verb?                      - DF(al~,phi,phi~))?
\par\noindent
Evidently, it makes sense to
\dfiline\verb?  resolve EQUATION (1) w.r.t. df(h,y,2);?%
\par\noindent
 The result is \EqN{2} (we shall display it explicitly below) while the
preceding \vb{EQUATION}
may be removed (by means of the instruction \Vb{erase
EQUATION (1)}).

 The last calculation we carry out in frames of the current step is
the deriving of a consequence of \EqN{2} (which determines
\vb{df(h,y,2)}, see below) by means of its differentiating with
respect to the \vb{COORDINATE} \vb{y}. Specifically, let us claim to
\dfiline%
\verb?isolate EQUATION DR(LHS(EQ(2))-RHS(EQ(2)),y)=0;?
\dsiline\verb?show EQUATIONS;?
\par\noindent
\Rmr{Here the symbol
 \Vb{DR} is so called {\em crossderivative\/} operator which is
similar the to partial derivative but takes into account the mutual
tree-like dependences of the \vb{SCALARS}, if any, regarding them as
functions, not mere variables. Mathematically, one may regard
\vb{DR(}\dots\vb{,y)} in the above instruction as a partial
derivative with respect to \vb{y}, provided a proper interpretation
of dependences of variables (\vb{SCALARS} and \vb{COORDINATES}) is
implied.%
    }
This calculation yield a new \vb{EQUATION} and
we obtain in particular the following output:
\dfoline\verb? Eq. (3) reads DF(h,y,3)=0?
\someskip
\dsoline\verb? Eq. (2) reads DF(h,y,2)?
\someskip
\dsoline\verb?                   - al - al~?\npb
\dsoline\verb?                =E           *(DF(al,phi,phi~) + DF(al~,phi,phi~))?
\par\noindent

 The current point is just reasonable for the terminating of the present
(the first) step of the problem solving. The result having been
derived is the following: we determined the dependence of all the
unknowns on the \vb{COORDINATE} \vb{y} (which represents the
affine parameter $y$ along the multiple null principal congruence).
Specifically, the \vb{SCALARS}
\vb{al}, \vb{chi} are independent on it (as well as \vb{al\ovrl},
\vb{chi\ovrl}) while \vb{h} is a second order polynomial in \vb{y},
the higher order coefficient being found in explicit form.

Additionally,
we have seen that
a gauge transformation allows to add any function of
$x,\phi,\bar\phi$ to
$\chi\strut,_{\bar\phi}
+
\bar\chi\strut,_\phi$.
See Eq.\ (\ref{chigau}).
 Thus since both $\chi$ and $\bar\chi$ are independent on $y$ the
above 2-term sum can be nullified by means of a proper gauge choice.
Then, there is a real `potential' $k=k(x,\phi,\bar\phi)$ such that
$\chi=\imu\,k\strut,_\phi, \bar\chi=-\imu\,k\strut,_{\bar\phi}$.
Ultimately, noting that the functions $\chi, \bar\chi$ enter the
tetrad in the form of the expression $\Re(\chi\,\dex\,\phi)$ alone,
we have the reduced representation
 \begin{equation}
 \Re(\chi\,\dex\,\phi)=-\Im(k\strut,_\phi\,\dex\,\phi)
                                        \label{chik}
\end{equation}
which will be used below in `formulae' (\Grg\ scripts)
determining \vb{TETRAD}.

 An additional remark is now in order. It has to be mentioned that
in this section we exhibited, essentially, the complete
session of the work with \Grg. Indeed, the input code used in the
first step of calculations consists, at first, of the section of \vb{DATA}
which was displayed in the beginning and, at second, a series of
instructions (assumed here to be issued from a keyboard) which was
listed (and commented) in the course of the discussion as well%
 \footnote{To be completely precise, the file {\sf slang} containing
    the additional
    section of \vb{CONFIDENTIAL SLANG} which installs the
    extended collection of `keywords' is read and included in the
    problem specification (\cf\ the list line of the script displayed
    in Appendix A). We omit discussion of this facility here
    however. See \cite{Grg}.
           }.
We also
reproduced all the substantial output which follow the instructions
execution (the only omission is the initialization messages and the
outcome of some preliminary analysis of the initial data by the
system including the copy of the structured problem specification).

 Unfortunately, in the sequel, in many cases the authentic copies of
output would occupy too much space to be given here. For example, the
output of the only next (the second) step of calculation
considered below is more
that 30 kB long, \ie, would require at least 10 pages even in a
condensed format. Hence we shall carry out below a strict selecting
and shall display only those output which is actually necessary. On
the contrary, the input will be displayed without omissions.

\subsection{Step 2: Introduction of electromagnetic field
and further reduction of field equations}

The \vb{DATA} incorporated in the problem
specification of the second step of calculations
develop ones used in the preceding step
(see above lines {\fn{1}--\fn{11}}).
 In particular, the symbols of \vb{COORDINATES}
need not be varied, of course. Similarly, the \vb{TETRAD} is described by
the same {\em paragraph\/} (a fragment of a section bounded at
right by the semicolon \Vb{;}) which comprises lines
{\fn{6}--\fn{10}}), where however the equation (\ref{chik}) is
taken into account. Thus now instead of the line {\fn{10}}, one has
the following one:
\filine{10n}\verb?    component|3=(d y -2 IM(df(k,phi) d phi)?%
\verb? +h d x)/sqrt(2);?
\par\noindent
Furthermore, now the identifier \vb{h} is not an
abstract `\vb{SCALAR}'. Instead, it is specified an \vb{ABBREVIATION}
by means of the statement
\filine{0h}\verb?  ABBREVIATION follows: h=R+Q*y+P*y**2;?
\par\noindent
\Rmr{
 An identifier of \vb{ABBREVIATION} is immediately replaced by the
corresponding expression associated with it whenever it is met in a
mathematical formula}.
 Thus we realize the form of the dependence of $h$ upon $y$ (second
order polynomial) inferred in the preceding section from the field
equations. The thee symbols \vb{R},\vb{P},\vb{Q} are the new
\vb{SCALARS} instead of the single \vb{h} but, as opposed to \vb{h},
they are independent on \vb{y}. We shall need even more
\vb{SCALARS}. Their total collection looks now as follows:
\filine{1s}\verb?  declare SCALARS al,k,R,P,Q(x,phi,phi~),?
\szline\verb?                  gamma,tau,zeta,lam,mu(x,y,phi,phi~),?
\siline{3s}\verb?                  psi,ems(x,y,phi,phi~);?
\par\noindent
(\cf\ line {\fn{3}}).
Besides, let us
\filine{0re}\verb?  declare REAL x,y,k,h,R,P,Q;?
\dsiline%
\verb?  Declare COMPLEX CONJUGATED phi & phi~,al & al~,lam & lam~,?
\dsiline\verb?                             mu & mu~,zeta & zeta~,?
\dsiline\verb?                             psi & psi~, ems & ems~;?
\par\noindent
Further, we may specify the structure of \vb{UNDOTTED CONNECTION}
basing on its properties derived in the preceding
subsection. Specifically,
\dfline\verb?  UNDOTTED CONNECTION comprises?
\siline{44}\verb?   component|1=gamma d phi + tau d phi~ + zeta d x,?
\siline{45}\verb?   component|2=(mu d phi~ + lam d x)/E**al;?
\par\noindent
\Rmr{
 Dropping out specification of the \vb{component\vrt
0} of \vb{UNDOTTED CONNECTION}
($\mGG{0}$ in the mathematical notation) in this {\em working
binding\/} of the data object value, it is automatically nullified.%
 }
 This representation of connection components merely expresses the
form of their expansions implied by the equations displayed in the
lines {\fn{10g}--\fn{11g}}, {\fn{20g}--\fn{21g}}, see the comment
following them.
The denominator
\vb{E**al} (\ie\ $\eul^\alpha$) of the \vb{component\vrt 2}
is introduced for the sake of the later
convenience (it makes some relationships simpler).
The \vb{SCALARS} \vb{gamma}, \vb{tau}, \vb{zeta}, \vb{mu}, \vb{lam} are
assumed to be subjected no \apriori\ constraints, being therefore
defined just by the expansions displayed in lines {\fn{44}--\fn{45}}.

Accomplishing the \vb{DATA} section of the current version of the
problem specification, we add the following description of the
electromagnetic field:
\dfiline\verb?  UNDOTTED EM SPINOR components are?
\siline{50}\verb?     component|1=psi,?
\siline{51}\verb?     component|2=ems;?
\siline{52}%
\verb?  DOTTED EM SPINOR is HERMITEAN CONJUGATED ?
\verb?to UNDOTTED EM SPINOR;?
\par\noindent
Since the \vb{SCALARS} \vb{psi}, \vb{ems} still correspond to
arbitrary functions (see line {\fn{3s}}),
it, essentially, only states that the
\vb{component\vrt0} (encoding $\mbEEM{0}=\mbEM{00}$)
 vanishes, implying additionally (by
virtue of the statement displayed line {\fn{52}}) that $\mdbEEM{0}$
vanishes as well.
 It is worth noting
that the vanishing of $\mbEEM{0}$ ($\mdbEEM{0}$) is
neither a restriction nor an ansatz since it can be deduced from
Eqs.~(\ref{Om0}), (\ref{raev}) and the spinor part of
Einstein--Maxwell equations (\ref{spp}) in a way similar to one used
for the derivation of Eqs.~(\ref{III}).

\smallskip

 Proceeding now with calculations, let us first of all examine
implications of the first structural equations. Since, given the
tetrad, they have to uniquely determine the connection forms, it
may be expected that, in the case under consideration, the explicit
values of the \vb{SCALARS} \vb{gamma}, \vb{tau}, \vb{zeta}, \vb{mu},
\vb{lam} have to follow.
 This is indeed the case.
Namely, the instruction
\dfiline\verb?obtain and isolate?
\dsiline\verb?      the SECOND VERSION of the UNDOTTED FIRST CARTAN EQUATIONS?
\par\noindent
entails just the five \vb{EQUATIONS}
(automatically enumerated with the numbers \vb{1},\dots,\vb{5}).
Then, requiring to
\filine{5k}\verb?resolve EQUATIONS (1)-(5) w.r.t. gamma,tau,mu,zeta,lam?
\par\noindent
(after that one may \Vb{erase EQUATIONS (1)-(5)})
and then, issuing the instructions
\filine{6k}\verb?turn on the displaying of negative powers;?
\siline{7k}\verb?show new EQUATIONS factoring out E**al,E**al~?
\par\noindent
we obtain the following output:
\dfoline\verb?                           1?\npb
\dsoline\verb?  Eq. (10) reads gamma= - ---*DF(al~,phi)?\npb
\dsoline\verb?                           2?
\someskip
\dsoline\verb?                     1?\npb
\dsoline\verb?  Eq. (9) reads tau=---*DF(al,phi~)?\npb
\dsoline\verb?                     2?
\someskip
\dsoline\verb?                        1   al + al~?\npb
\dsoline\verb?  Eq. (8) reads mu= - (---*E        )*(DF(al,x) + DF(al~,x))?\npb
\dsoline\verb?                        2?
\someskip
\dsoline\verb?                    - DF(k,phi,phi~)*I?
\someskip
\soline{1zt}\verb?                      1              1                     1?\npb
\sozline\verb?  Eq. (7) reads zeta=---*DF(al,x) - ---*DF(al~,x) + P*y + ---*Q?\npb
\sozline\verb?                      4              4                     2?
\someskip
\sozline\verb?                    1    al~ -1   al -1?\npb
\sozline\verb?                 + ---*(E   )  *(E  )  *DF(k,phi,phi~)*I?\npb
\soline{6zt}\verb?                    2?
\someskip
\dsoline\verb?  Eq. (6) reads lam= - DF(k,x,phi)*I - 2*DF(k,phi)*I*P*y?
\someskip
\dsoline\verb?                                                              2?\npb
\dsoline\verb?                     - DF(k,phi)*I*Q + DF(R,phi) + DF(P,phi)*y?
\someskip
\dsoline\verb?                     + DF(Q,phi)*y?
\par\noindent
These relationships are just the content of the first structural equations.
Essentially, they define the connection forms, see lines \fn{44}--\fn{45}.

 We may regard the expressions determining auxiliary \vb{SCALARS}
\vb{gamma}, \vb{tau} as the ultimate formulae and use them for the
complete eliminating these \vb{SCALARS} (which are involved in the
expansions of \vb{UNDOTTED CONNECTION} alone). To that end, we have
preliminarily to
\dfiline\verb?add substitution rules LHS(EQ(9))->RHS(EQ(9)),?
\dsiline\verb?                       LHS(EQ(10))->RHS(EQ(10))?
\par\noindent
 to the section of \vb{SUBSTITUTIONS}. (Although we have not
included this section it in the problem specification, it is automatically
introduced.)
 \Rmr{The above record of substitution rules mean that,
bringing them into action, the
expression at left in \vb{EQ}[\vb{.}]\vb{(9)} displayed above
is replaced by the expression situated at right (which is indicated
by the symbol \Vb{->}); the second formula involving
\vb{EQ}[\vb{.}]\vb{(10)} possesses similar meaning.}

The
substitutions introduced are automatically endowed with the numbers
\vb{1} and \vb{2}, respectively%
 \footnote{One might introduce any other numerical
  labels instead.}%
.
They are used for the referring to the rules.
It particular, we invoke the substitutions which ensure the eliminating of
\vb{gamma} and \vb{tau} in favor of their values by means of the instruction
\dfiline\verb?match the rules (1),(2) with UNDOTTED CONNECTION?
\par\noindent
(In order to
verify that the necessary transformation has been indeed carried out
 one may issue the instruction \Vb{type UNDOTTED CONNECTION}.)

 Next we consider a component of Einstein field equations which was
also used in the first step of calculation. The instruction
\dfiline\verb?obtain and isolate the SCALAR PART of EINSTEIN-MAXWELL EQUATIONS?
\par\noindent
entails the only new \vb{EQUATION} endowed with the number \vb{11}.
Requiring to \Vb{show new EQUATION}, we obtain in particular the output
\dfoline\verb?                      al + al~?\npb
\dsoline\verb?  Eq. (11) reads 0=2*E        *DF(zeta,y) - DF(al,phi,phi~)?
\someskip
\dsoline\verb?                    - DF(al~,phi,phi~)?
\par\noindent
 Notice that this \vb{EQUATION} involves the derivative of the
\vb{SCALAR zeta} whose value has been specified by the above
\EqN{7}. A natural consequence of these two \vb{EQUATIONS},
implying the elimination of \vb{DF(zeta,y)}, may be derived by means
of the instruction
\dfiline\verb?isolate EQUATION LHS(EQ(11))=RHS(EQ(11))?
\dsiline\verb?  -COEFFN(RHS(EQ(11)),df(zeta,y))*DR(LHS(EQ(7))-RHS(EQ(7)),y)?
\par\noindent
 This is an example of operation frequently used for reductions of
overdetermined systems of equations with partial derivatives.
Removing the second line we
would obtain the \vb{EQUATION} identical to the initial one (number
\vb{11}). The additional term in the lower line is proportional to
the derivative of the {\it discrepancy\/} \Vb{LHS(EQ(7))-RHS(EQ(7))}
of \EqN{7}. Thus the resulting \vb{EQUATION} (number \vb{12}) is
equivalent to \EqN{11} {\it modulo\/} \EqN{7}. The coefficient in
front of the derivative of the discrepancy is chosen in such a way to
automatically entail the desirable simplification. In our case we have to
eliminate \vb{df(zeta,y)}; hence it equals
\vb{COEFFN(RHS(EQ(11)),df(zeta,y))}, \ie\ the coefficient in front
of the derivative \vb{df(zeta,y)} involved in r.h.s.\ expression of
\EqN{11} which is considered here as polynomial in \vb{df(zeta,y)} (a
linear function in fact).

It turns out that the last \vb{EQUATION} obtained determines the value of
the \vb{SCALAR} \vb{P}. The instructions
\dfiline\verb?resolve Eq. (12) w.r.t. P; show new EQUATIONS?
\par\noindent
yield in particular
\foline{1P}\verb?                    1    - al - al~?\npb
\sozline\verb?  Eq. (13) reads P=---*E?\npb
\sozline\verb?                    2?
\someskip
\soline{4P}\verb?                   *(DF(al,phi,phi~) + DF(al~,phi,phi~))?
\par\noindent
which will play an important role in what follows.

Now the optimal way of the further reduction of the field equation
is to proceed with Maxwell equations. Thus the next instruction reads:
\dfiline\verb?obtain and isolate UNDOTTED MAXWELL EQUATIONS?
\par\noindent
 It yields four `plain' \vb{EQUATIONS} (number \vb{14}-\vb{17})
which can be found to determine a number of \vb{SCALAR} derivatives.
Specifically, the instructions
\filine{20q}\verb?resolve Eqs. (14)-(17) w.r.t.?
\siline{21q}\verb?             DF(psi,y),DF(ems,y),DF(psi,phi~),DF(psi,phi);?
\dsiline\verb?show new EQUATIONS?
\par\noindent
yield the corresponding explicit expressions (\vb{Eqs.\ (18)}-\vb{(21)}).
We obtain in particular that
\foline{1p}\verb?  Eq. (21) reads DF(psi,y)=0?
\par\noindent
and
\foline{8p}\verb?  Eq. (19) reads DF(psi,phi~)=0?
\par\noindent
(We do not display the values of \vb{DF(ems,y)} (\EqN{20})
and
\vb{DF(psi,phi)}  (\EqN{18})
since it is only essential currently to know
that these are nonzero).
 We see therefore that the function $\psi$ (represented by the
\vb{SCALAR psi}) which equals the component $\mbEEM{1}$ of the
undotted spinor of electromagnetic field (see line \fn{50}) depends
on the coordinates $x$ and $\phi$ alone (represented by the symbols
\vb{x}, \vb{phi}). Since $\phi$ is a complex variable $\psi$ has to
be holomorphic with respect to it (and is assumed to be $C^{\infty}$
with respect to $x$). Further, we have to inform the algebraic
processor that the constraints \fn{1p},\fn{8p} are to be taken into
account throughout all the subsequent calculations. This is realized
by means of the instruction
\dfiline\verb?let the last EQUATION hold true,?
\dsiline\verb?    the last but 2 EQUATION hold true?

 A series of the further manipulations with the collection of
\vb{EQUATIONS} obtained so far follows a standard routine procedure
which widely applies for the reducing of overdetermined systems of
quasilinear equations with partial derivatives%
 \footnote{For the first reading it might be recommended to
  skip the fragment below up to the line \fn{2y}.}.
Specifically, some
of the equations available are differentiated and, then, combined in
such a way to cancel out the higher order derivatives arisen. Such
an operation is repeated until it yields new tractable
relationships. Thus, following this scheme, the concrete
transformations, which lead to the desirable simplifications and
entail new data (new \vb{EQUATIONS}), are chosen by ourselves with
the help of the analysis of the intermediate output. In all the
cases we meet the latter is fairly straightforward but,
nevertheless, too bulky to be displayed here. Hence we shall not
exhibit all these speculations and shall outline only the most
substantial points of the derivation whose code is displayed below.

Specifically, the following series of instructions is issued:
\filine{42q}\verb?isolate EQUATION DR(LHS(EQ(18)),y)=DR(RHS(EQ(18)),y);?
\siline{43q}\verb?show new EQUATIONS;?
\siline{44q}\verb?resolve Eqs. (20),(23) w.r.t. DF(ems,y),DF(ems,y,phi~);?
\siline{45q}\verb?show new EQUATION;?
\siline{47q}\verb?let (10) the last EQUATION hold true;?
\siline{48q}\verb?isolate EQUATION DR(LHS(EQ(20)),y)=DR(RHS(EQ(20)),y);?
\siline{49q}\verb?resolve Eqs. (25),(20) w.r.t. DF(ems,y,2),ems;?
\siline{50q}\verb?abolish substitution (10);?
\dsiline\verb?show Eq. (27);?
\siline{53q}\verb?let the last EQUATION hold true;?
\siline{54q}\verb?resolve Eqs. (18),(26) w.r.t. DF(k,phi,phi~),df(ems,y);?
\dsiline\verb?show new EQUATIONS;?
\siline{58q}\verb?resolve Eqs. (26),(28) w.r.t. df(ems,y),ems;?
\dsiline\verb?show new EQUATION;?
 \par\noindent
\EqN{18} (produced by the instruction displayed in lines \fn{20q}-\fn{21q})
expresses \vb{DF(psi,phi)}.
 By virtue of \EqN{21} {\sc l.h.s.} of the \vb{EQUATION} described
in the line \fn{42q} vanishes. Explicitly, the corresponding new
\dfiline\verb? Eq. (23) reads 0=DF(al,phi~)*DF(ems,y) + DF(ems,y,phi~)?
 \par\noindent
(a part of the output of line \fn{43q}).
 In turn, \EqN{20} expresses \vb{DF(ems,y)} and lines
\fn{44q}-\fn{45q} yield separate representations for
\vb{DF(ems,y,phi\ovrl)} and \vb{DF(ems,y)}. The instruction yields a
single new \vb{EQUATION} (expressing \vb{DF(ems,y,phi\ovrl)}%
  \footnote{Representation of \vb{DF(ems,y)} does not vary; its
inclusion to the instruction \fn{44q}-\fn{45q} is intended for the
{\em eliminating\/} of \vb{DF(ems,y)}. Such a
trick will be often used below.}%
  ).
 Line \fn{47q} initiates subsequent replacing of the derivative
\vb{DF(ems,y,phi\ovrl)} by the corresponding value, the rule being
endowed with number \vb{10}. (In line \fn{50q} this substitution is
disabled.) \EqN{20} determines \vb{DF(ems,y)}. Thus in the line \fn{48q}
the second order derivative \vb{DF(ems,y,2)} is introduced.

The
instruction in line \fn{49q} eliminates \vb{ems} from {\sc r.h.s.}
of the new \vb{EQUATION} (number \vb{25}). One of the resulting
\vb{EQUATIONS} is most important, namely,
\foline{2y}\verb?  Eq. (27) reads DF(ems,y,2)=0?
 \par\noindent
Thus \vb{ems} is linear in \vb{y}.
Line \fn{53q} ensures the nullifying of \vb{DF(ems,y,2)}
in all the subsequent calculations.

It is straightforward now to determine the coefficient in front of \vb{y}
in \vb{ems} expansion.
The instruction in line
\fn{54q} yields a pair of \vb{EQUATIONS}. In particular
\foline{28a}\verb?                           - al?\npb
\soline{28b}\verb? Eq. (28) reads DF(ems,y)=E     *DF(psi,phi)?
 \par\noindent
(\vb{psi} does not depends on \vb{y}).
 Finally, line \fn{58q} yields representation of \vb{ems} through
\vb{DF(ems,phi\ovrl)} and the expressions independent on \vb{y}
(\EqN{30}) which will be utilized later on.

\smallskip

Next, we shall deduce some useful consequences of Einstein--Maxwell
equations.
To that end, let us, at first,
\dfiline\verb?obtain the SPINOR PART of EINSTEIN-MAXWELL EQUATIONS;?
\dsiline\verb?isolate the component|1|1~ of the ABOVE EQUATIONS;?
\dsiline\verb?show new EQUATIONS?
\par\noindent
This yields in particular the output:
\dfoline\verb?                         al + al~               al + al~?\npb
\dsoline\verb?  Eq. (31) reads 0= - 2*E        *DF(zeta,y) - E        *psi*psi~?
\someskip
\dsoline\verb?                    - DF(al,phi,phi~) - DF(al~,phi,phi~)?
\par\noindent
 \EqN{31} involves the derivative \vb(zeta,y). On the other hand we
had the `explicit' representation of \vb{zeta} (\EqN{7}, see lines
\fn{1zt}-\fn{6zt}). The dependence of \vb{zeta} on \vb{y} is
characterized by the \vb{SCALAR} \vb{P} which, in turn, is
determined by \EqN{13} (lines \fn{1P}-\fn{4P}). These relationships
(determining \vb{DF(zeta,y)}) can be represented by means of
the two substitution rules which are introduced as follows:
\dfiline\verb?add substitution rules?
\dsiline$\!$\verb? (30) P -> P-(LHS(EQ(13))-RHS(EQ(13))),?
\dsiline$\!$\verb? (31) DF(zeta,y) -> ?%
\verb?DF(zeta,y)-MATCHING(DR(LHS(EQ(7))-RHS(EQ(7)),y),30)?
\par\noindent
Notice that,
 formally, the rule \vb{(30)} is trivial {\it modulo\/} \EqN{13}.
However, the term involving it is chosen in such a way to eliminate
\vb{P} at right.
Similarly, the rule \vb{(31)} is trivial
{\it modulo\/} some expression vanishing if \EqN{7} is satisfied.
\Rmr{Here the {\em macro\/} \vb{MATCHING} applies the substitution rule
whose number is given as the second argument%
 \footnote{An arbitrary number the rules is allowed to be invoked by the
           \Vb{MATCHING} operator.}
(\ie, the rule \vb{(30)}) to the first argument.}
The combined effect is the eliminating of the both
\vb{DF(zeta,y)} and \vb{P} at right in the expression to be
substituted in accordance with the rule \vb{(31)}.
Accordingly, applying it to \EqN{31} by means of the instruction
\dfiline\verb?isolate EQUATION 0=MATCHING(LHS(EQ(31))-RHS(EQ(31)),31);?
\par\noindent
(here \Vb{31} in \vb{EQ(31)} is the number of an \vb{EQUATION}
while the last \Vb{31} is the number of a {\em substitution rule};
their coincidence is occasional),
the request \Vb{show new EQUATION}
yields in particular the following output:
\foline{1li}\verb?                    al + al~?\npb
\sozline\verb?  Eq. (32) reads 0=E        *psi*psi~ + 2*DF(al,phi,phi~)?
\someskip
\soline{3li}\verb?                    + 2*DF(al~,phi,phi~)?
\par\noindent

 This \vb{EQUATION} is worth a separate attention. It involves the
\vb{SCALARS} \vb{al}, \vb{al\ovrl}, \vb{psi}, \vb{psi\ovrl}
(representing $\alpha, \bar\alpha, \psi, \bar\psi$) which may be
regarded as functions of some fixed sets of arguments. In accordance
with declaration shown in lines {\fn{1s}--\fn{3s}}, \vb{al} and
\vb{al\ovrl} depend on \vb{x}, \vb{phi}, and \vb{phi\ovrl} while
more strong restriction is imposed on \vb{psi} (see lines {\fn{1p}},
{\fn{8p}}): it may depend on \vb{x} and \vb{phi} alone (\ie, is
holomorphic with respect to \vb{phi}). Then, since \vb{psi} and
\vb{psi\ovrl} are \vb{COMPLEX CONJUGATED}, the \vb{SCALAR}
\vb{psi\ovrl} depends on \vb{x} and \vb{phi\ovrl} (and represents
the function of $x$ and $\bar\phi$ holomorphic with respect to the
second argument). Taking these relationships into account, we see
that \EqN{28} may be regarded as {\em Lioville equation\/} with
respect to unknown \vb{al+al\ovrl} ($2\re\alpha$) for some given
\vb{psi}, \vb{psi\ovrl} ($\psi$ and $\bar\psi$), holomorphic and
antiholomorphic, respectively
 (\cf\ \cite{Cat}, Eq.\ (27.50)).
Besides, we have seen above that the value of
\vb{al-al\ovrl} is a matter of a gauge, provided it does not depend on
the \vb{COORDINATE} \vb{y}. Other relevant gauge freedom involves
the transformation of the \vb{COORDINATE} \vb{phi}
(and, simultaneously, its complex conjugated \vb{phi\ovrl})
which, in
mathematical notations, is described by the formula
$\phi\rightarrow\Xi(\phi,x)$ for arbitrary $\Xi$ holomorphic with
respect to $\phi$,
assuming the restriction
$\Xi(\phi,x),_\phi\ne 0$ to be fulfilled.
The latter fact has to be taken into account when analyzing solutions
of Lioville equation and their gauge transformations.

It should be noted now that
\Grg\ is not able (and was not intended) to solve differential
equations%
 \footnote{This point should not be considered as its inalienable
feature. Moreover, it could be worth introducing an interface with an
appropriate ODE or PDE package. However such a facility has not been
realized so far.}.
 We have to perform this work ourselves. In our case it is
straightforward to find and write down the general solution to
\EqN{32} (\cf\ \cite{Cat}, Eq.\ (27.51)). It is
convenient to represent the solution in the form of substitution
rules. They are introduced by the following instruction:
\dfiline\verb?add rules?
\siline{7s}\verb?  (40) al  -> -log(psi)-log(1+phi*phi~/4),?
\szline\verb?  (41) al~ -> -log(psi~)-log(1+phi*phi~/4),?
\szline%
\verb?  (42) df(al,phi,phi~) -> DR(-log(psi)-log(1+phi*phi~/4)?
\verb?,phi,phi~),?
\siline{10s}%
\verb?  (43) df(al~,phi,phi~) -> DR(-log(psi~)-log(1+phi*phi~/4)?
\verb?,phi,phi~);?
\par\noindent
 (here, evidently, the third and the fourth rules
are the consequences of the first and second ones) which
just describe the general local solution to \EqN{32}).

It is instructive to show that we indeed deal with a solution
the Lioville equation (\EqN{32}). Preliminarily, we have to inform the
system on the properties of the \vb{SCALAR} \vb{psi\ovrl} (as opposed
to \vb{psi}, we have not dealt with it yet in fact). It is convenient to
do this referring to the corresponding \vb{EQUATIONS} concerning its complex
conjugated counterpart, \vb{psi}, \ie\ \EqN{19} and \EqN{21}.
Taking their complex conjugation we obtain the appropriate
characteristics of \vb{psi\ovrl}.
To that
end, let us issue the instructions
\dfiline\verb?evaluate aux=C.C.(LHS(EQ(19)));?
\dsiline\verb?let (50) aux->C.C.(RHS(EQ(19)));?
\dsiline\verb?evaluate aux=C.C.(LHS(EQ(21)));?
\dsiline\verb?let aux->C.C.(RHS(EQ(21)));?
\par\noindent
 \Rmr{Here and below the (`user-defined') identifier \vb{aux}
plays role of auxiliary variable used for the temporary storing the
expressions situated on {\sc r.h.s.}\ of assignments in instructions
with the action \vb{EVALUATE}.}
   These substitution rules are labeled with
numerical labels \vb{50} (specified by ourselves) and \vb{51}
(the next number generated automatically).

Then the instruction
\filine{28s}\verb?evaluate aux=MATCHING(LHS(EQ(32))-RHS(EQ(32)),42,43),?
\siline{30s}\verb?         MATCHING(aux,40,41);?
\par\noindent
reports zero value which just means the expectable satisfaction of the \EqN{32}
(see lines
{\fn{1li}--\fn{3li}}) by virtue of the relationships
displayed in lines {\fn{7s}--\fn{10s}}.
(Let us
remind that lines {\fn{28s},\fn{30s}} re-evaluate the
`discrepancy'
of \EqN{32} executing, at first, substitutions number \vb{42},\vb{43},
and then, substitutions number \vb{40},\vb{41} which were
specified in lines {\fn{7s}--\fn{10s}}).

An additional useful result immediately follows from
\EqN{13} and \EqN{32}. Their appropriate superposition is realized
by means of instructions
\dfiline\verb?   isolate EQUATION?
\dsiline\verb?         LHS(EQ(9))=RHS(EQ(9))?
\dsiline{\hfuzz=1.9pt\relax%
\verb?                    -COEFFN(NUMR(RHS(EQ(9))),DF(al,phi,phi~))?
\dsiline\verb?                     /DENM(RHS(EQ(9)))/2?}
\dsiline\verb?                     *(LHS(EQ(280))-RHS(EQ(28)));?
\siline{31e}\verb?   show equation (29);?
\dfiline\verb?isolate EQUATION?
\dsiline\verb?   LHS(EQ(13))=RHS(EQ(13))?
\dsiline\verb?               +(COEFFN(NUMR(RHS(EQ(13))),DF(al,phi,phi~))?
\dsiline\verb?                +COEFFN(NUMR(RHS(EQ(13))),DF(al~,phi,phi~)))?
\dsiline\verb?                 /DENM(RHS(EQ(13)))?
\dsiline\verb?                *(LHS(EQ(32))-RHS(EQ(32)))?
\dsiline\verb?                 /(COEFFN(NUMR(RHS(EQ(32))),DF(al,phi,phi~))?
\dsiline\verb?                  +COEFFN(NUMR(RHS(EQ(32))),DF(al~,phi,phi~)));?
\dsiline\verb?   show new EQUATION;?
\par\noindent
In particular, the last line entails the the response
\foline{P1}\verb?                       1?\npb
\sozline\verb?  Eq. (33) reads P= - ---*psi*psi~?\npb
\soline{P3}\verb?                       4?
\par\noindent

\smallskip

 The last action we shall carry out in frames of the current step is
the explicit separating of the dependence of the \vb{SCALAR} \vb{ems} on
the \vb{COORDINATE y}. As we have seen it is a linear function of
\vb{y} (see line {\fn{2y}}). Moreover, the coefficient in front of
\vb{y} is specified by \EqN{28} (lines
{\fn{28a}--\fn{28b}}). Thus it retains to find the free term
which
can be explicitly represented as \vb{ems-y*df(ems,y)}.
Instead we shall calculate the derivative of \vb{(ems-y*df(ems,y))*E**al}
with respect to \vb{phi\ovrl}. The corresponding instructions are
straightforward, executing several transformations which are
identical ones \mod\ \EqN{28} and \EqN{30}:
\dfiline\verb?   evaluate?
\siline{1exp}\verb?        expr=(ems-y*df(ems,y))*E**al?
\siline{2exp}\verb?        aux=DR(expr,phi~),?
\dsiline%
\verb?        aux=aux-COEFFN(NUMR(aux),df(ems,phi~,y))/DENM(aux)?
\dsiline\verb?                *DR(LHS(EQ(28))-RHS(EQ(28)),phi~),?
\dsiline\verb?        aux=aux-COEFFN(NUMR(aux),df(ems,y))/DENM(aux)?
\dsiline\verb?                *(LHS(EQ(28))-RHS(EQ(28))),?
\dsiline\verb?        aux=aux-COEFFN(NUMR(aux),ems)/DENM(aux)?
\dsiline\verb?                *(LHS(EQ(30))-RHS(EQ(30))),?
\dsiline\verb? aux?
\dsiline\verb?     factoring out E**al,E**al~;?
\par\noindent
 Here the line \fn{1exp} plays role of the definition of \vb{expr}
to be estimated, line \fn{2exp} introduces its derivative denoting it \vb{aux}
which is further undergone
transformations `identical' {\it modulo\/} \EqN{28}, \EqN{30}.
The last two lines of the code
exhibits its resulting representation:
\dfoline\verb?  The expression?
\dsoline\verb?      aux?
\dsoline\verb?                  amounts to:?
\someskip
\dsoline\verb?         al + al~?\npb
\dsoline\verb?      - E        *(DF(al,x)*psi + DF(psi,x) + DF(al~,x)*psi)?
\someskip
\dsoline\verb?         -1?\npb
\dsoline\verb?      + I  *(2*DF(k,phi,phi~)*psi + DF(k,phi~)*DF(psi,phi))?\rzr%
\verb?1.3 s.?
\par\noindent
The dependence derived
can be partially integrated by means of
the extracting from the (value of) auxiliary variable \vb{expr} a
series of appropriate expressions which, after the subtracting their
corresponding derivatives from the value of \vb{aux} (equal to
\vb{DR(expr,phi\ovrl)} \mod\ \EqN{28} and \EqN{30}), lead to the
cancelling out certain terms of the latter. Specifically, the
following instructions are to be issued:
\dfiline\verb?evaluate?
\siline{0hi}\verb?  expr=expr -df(k,phi)* ?
\verb?COEFFN(NUMR(aux),df(k,phi,phi~))/DENM(aux),?
\szline\verb?  aux=DR(expr,phi~),?
\szline\verb?  aux=aux-COEFFN(NUMR(aux),df(ems,phi~,y))/DENM(aux)?
\szline\verb?          *DR(LHS(EQ(28))-RHS(EQ(28)),phi~),?
\szline\verb?  aux=aux-COEFFN(NUMR(aux),df(ems,y))/DENM(aux)?
\szline\verb?          *(LHS(EQ(28))-RHS(EQ(28))),?
\szline\verb?  aux=aux-COEFFN(NUMR(aux),df(k,phi,phi~))/DENM(aux)?
\siline{9hi}\verb?          *(LHS(EQ(29))-RHS(EQ(29))),?
\dsiline\verb?  aux;?
\someskip
\dsiline\verb?evaluate?
\siline{0hj}\verb?  expr=expr -k* COEFFN(NUMR(aux),df(k,phi~))/DENM(aux),?
\szline\verb?  aux=DR(expr,phi~),?
\szline\verb?  aux=aux-COEFFN(NUMR(aux),df(ems,phi~,y))/DENM(aux)?
\szline\verb?          *DR(LHS(EQ(28))-RHS(EQ(28)),phi~),?
\szline\verb?  aux=aux-COEFFN(NUMR(aux),df(ems,y))/DENM(aux)?
\szline\verb?          *(LHS(EQ(28))-RHS(EQ(28))),?
\szline\verb?  aux=aux-COEFFN(NUMR(aux),df(k,phi,phi~))/DENM(aux)?
\siline{9hj}\verb?          *(LHS(EQ(29))-RHS(EQ(29))),?
\dsiline\verb?  aux;?
\someskip
\dsiline\verb?  evaluate  expr,aux-DR(-E**(al+al~)*psi,x)?
\dsiline\verb?            factoring out E**al;?
\par\noindent
Here the first application of the action \Vb{EVALUATE}
(lines {\fn{0hi}--\fn{9hi}})
`integrates out' the term proportional to \vb{df(k,phi)} while
the second application
(lines {\fn{0hj}--\fn{9hj}})
makes the same thing with the term proportional to \vb{k}.
The output of the last instruction looks as follows:
\dfoline\verb? The expression?
\dsoline\verb?     expr?
\dsoline\verb?                 amounts to:?
\someskip
\soline{1dex}\verb?     al?\npb
\sozline\verb?    E  *( - DF(ems,y)*y + ems)?
\someskip
\sozline\verb?        -1?\npb
\soline{3dex}\verb?     + I  *( - 2*DF(k,phi)*psi - DF(psi,phi)*k)?\rzr%
\verb?0.2 s.?
\someskip
\dsoline\verb? The expression?
\dsoline\verb?     aux - DR(- E** (al + al~) * psi, x)?
\dsoline\verb?                 vanishes.                     ?\rzr%
\verb?0.2 s.?
\par\noindent
Thus the derivative of the expression displayed in lines
{\fn{1dex}--\fn{3dex}} with respect to \vb{phi\ovrl} equals the
derivative of \verb?-E**(al+al~)*psi? with respect to \vb{x}.
Hence we obtain the free (\vb{y}-independent) term of \vb{ems}
by a quadrature.

Now the amount of the new information deduced suffices to renew
the problem specification.

 \subsection{Step 3: Accomplishing of reduction of the set of
             field equations}

We modify the preceding problem specification is such a way to
implement all the relationships obtained so far. We need now to
\dfiline\verb?  Declare SCALARS k,R,Q(x,phi,phi~),?
\siline{2sc}\verb?                  psi,beta(x,phi),psi~(x,phi~),?
\siline{3sc}\verb?                  xi(x,phi,phi~),?
\dsiline\verb?                  rho(phi,phi~),?
\dsiline\verb?                  lam(y,Q,psi,psi~,x,phi,phi~),?
\dsiline\verb?                  zeta(y,Q,rho,psi,psi~,x,phi,phi~),?
\dsiline\verb?                  mu(rho,psi,psi~,x,phi,phi~),?
\dsiline\verb?                  ems0(rho,psi,k,xi,x,phi,phi~),?
\dsiline\verb?                  ems1(rho,psi,x,phi);?
\par\noindent
 In addition to the \vb{SCALARS} used before and displayed in lines
{\fn{1s}--\fn{3s}} some new ones are introduced.
The collections of their `arguments'
are determined by the \vb{VALUES} which has to be bound with them.
The latter realize the relationships (equations) derived above.
Specifically,
\dfiline\verb?  SCALAR VALUES follow:?
\siline{0rho}\verb?        rho=1 +phi*phi~/4,?
\dsiline%
\verb?        lam=DR(P,phi)*y**2 +y*(df(Q,phi) -2i*P*df(k,phi))?
\dsiline\verb?            -i*df(k,x,phi) -i*Q*df(k,phi) +df(R,phi),?
\dsiline\verb?        zeta=P*y +DR(al-al~,x)/4?
\dsiline\verb?             +Q/2 +df(k,phi,phi~)*i/2/E**(al+al~),?
\dsiline\verb?        mu= -DR(E**(al+al~),x)/2 -i*df(k,phi,phi~),?
\dsiline\verb?        ems1=E**(-al)*df(psi,phi),?
\dsiline%
\verb?        ems0=E**(-al)*(-2i*psi*df(k,phi) -i*k*df(psi,phi)?
\dsiline\verb?                       -xi -df(beta,phi));?
\par\noindent
 As opposed to the interpretation followed to in the preceding
steps, the symbols \vb{al} (together with \vb{al\ovrl}) and \vb{P}
are now subsidiary objects.  They are defined in accordance with the
following statement:
\filine{1ab}\verb?  ABBREVIATIONS comprise?
\szline\verb?                  al=-log(rho)-log(psi),?
\szline\verb?                  P=-psi*psi~/4,?
\siline{4ab}\verb?                  h=R+Q*y+P*y**2;?
\par\noindent
\Cf\ lines \fn{7s}, \fn{P1}--\fn{P3}, \fn{0rho}.
 (Here \vb{h} definition exactly replicates the one used above, see
line {\fn{0h}}).

 The declaration of \vb{REAL} objects (see line {\fn{0re}}) is now to be
replenished with the \vb{SCALAR} \vb{rho} while all the other new
\vb{SCALARS} are defined as complex (by means of the declaration
\Vb{COMPLEX CONJUGATED}),
being endowed with complex
conjugated counterparts whose identifiers are
marked with the character \Vb{\ovrl}
attached at right.

 The specification of the basic geometric geometric objects ---
\vb{TETRAD} and \vb{UNDOTTED CONNECTION} --- consists of the same
formulae displayed in lines {\fn{6}--\fn{10}} and
{\fn{44}--\fn{45}}. As to the \vb{UNDOTTED EM SPINOR}, its
\vb{component|1} looks as above while the \vb{component|2} is endowed with
the expansion revealing its dependence on \vb{y}:
\dfiline\verb?         component|2=ems0+y*ems1;?
\par\noindent
 This line replaces line {\fn{51}}. In particular, the former \vb{SCALAR}
\vb{ems} (depending on \vb{y}) is now replaced by the two new ones,
\vb{ems0} and \vb{ems1}, which are independent on \vb{y}.

The last --- and novel --- element of the problem specification is the
following section of
\dfiline\verb? Substitutions:?
\dsiline\verb?  (1)  lam->VAL(lam);?
\dsiline\verb?  (2)  zeta->VAL(zeta);?
\dsiline\verb?  (3)  mu->VAL(mu);?
\siline{4sub}\verb?  (4)  rho->VAL(rho);?
\dsiline\verb?  (5)  ems1->VAL(ems1);?
\dsiline\verb?  (6)  ems0->VAL(ems0);?
\siline{7sub}\verb?  (7)  df(xi,phi~)->DR(e**(al+al~)*psi,x);?
\dsiline\verb?  (8)  ems1~->VAL(ems1~);?
\dsiline\verb?  (9)  ems0~->VAL(ems0~);?
\dsiline\verb? end of substitutions.?
\par\noindent
 These list of {\em substitution rules\/} will be utilized whenever
necessary. All they but one (displayed in line {\fn{7sub}}) mean the
replacing of the {\em identifier\/} of a \vb{SCALAR} by the
corresponding \vb{SCALAR VALUE} (`extracted' by the macros \vb{VAL}).
The specific substitution situated in line {\fn{7sub}} characterizes
\vb{xi} as the integral of the expression at right hand side with
respect to the variable (complex \vb{COORDINATE}) \vb{phi\ovrl}.

Now let us proceed with calculations.

At first, it is instructive to demonstrate how one can check
that the current initial data verifies the part of field equations
considered during the preceding steps of our calculation.
This is carried out by means of the following instructions:
\dfiline\verb?obtain the SECOND VERSION ?
\verb?of the UNDOTTED FIRST CARTAN EQUATIONS,?
\dsiline\verb?          UNDOTTED MAXWELL EQUATIONS,?
\dsiline%
\verb?      and the SCALAR PART of EINSTEIN-MAXWELL EQUATIONS;?
\siline{1mt}\verb?match substitution rules (1)-(4),(5)-(7) with?
\siline{2mt}\verb?      the ABOVE EQUATIONS;?
\siline{1ty}\verb?renew and type?
\dsiline\verb?      the ABOVE EQUATIONS;?
\par\noindent
 These equations are satisfied not automatically but as a
consequence of specific \vb{SCALAR VALUES} given in the problem
specification. Accordingly, their substituting realized by the
instruction displayed in lines \fn{1mt}-\fn{2mt} is necessary. The
output of the above {\em action\/} \vb{TYPE} situated in line
\fn{1ty} says us that all the equations  listed are
satisfied. Thus to obtain the `complete' solution of the problem it
retains to
\dfiline%
\verb?obtain the SPINOR PART of EINSTEIN-MAXWELL EQUATIONS?
\par\noindent
and ensure its fulfillment.

In this case,
similarly to the above procedure, we also have to take into account the
\vb{SCALAR VALUES} given (evaluating derivatives of \vb{SCALARS},
they are automatically taken into account). To that end, the
following instructions are to be issued:
\filine{0pd}\verb?match substitution rules (1)-(3),(5)-(6),(8)-(9) with?
\szline\verb?           the ABOVE EQUATIONS;?
\siline{4pd}\verb?isolate the ABOVE EQUATIONS;?
  \par\noindent
 The result of the line \fn{4pd} is four differential \vb{EQUATIONS}
(marked with the ordinal numbers \vb{1},\vb{2},\vb{3},\vb{4}).
The first of them
is in fact trivial, being immediately fulfilled, provided the
\vb{SCALAR VALUE} for \vb{rho} (see line {\fn{0rho}}) is taken into
account. Indeed, the instruction
\dfiline\verb?   Evaluate MATCHING(LHS(EQ(1))-RHS(EQ(1)),4);?
\par\noindent
reports zero result.
(The substitution number \vb{4} which is referred to here causes
the replacing of the \vb{SCALAR} \vb{rho} by the corresponding
\vb{VALUE}, see lines {\fn{4sub}},\fn{0rho}.)
On the contrary, a single \EqN{4} entails two more
\vb{EQUATIONS} (which prove to be non-trivial) since it means the
vanishing of a linear function of \vb{y}.
Accordingly, the instruction
\dfiline\verb?isolate?
\dsiline\verb?       EQUATION 0=COEFFN(LHS(EQ(4))-RHS(EQ(4)),y,1),?
\dsiline\verb?       EQUATION 0=COEFFN(LHS(EQ(4))-RHS(EQ(4)),y,0);?
\par\noindent
 constructs the \vb{EQUATIONS} (their ordinal numbers are \vb{5} and
\vb{6}) which just express the vanishing of the first (the first
line, the last parameter equals 1) and zero (the second line, the
last parameter equals 0) order coefficients.
 The resulting collection of four \vb{EQUATIONS}, which all the
field equations have been reduced to, can be resolved with respect
to derivatives of unknown \vb{SCALARS} and displayed by means of the
instructions
\dfiline\verb?resolve EQUATIONS (2),(3),(5),(6) w.r.t.?
\dsiline%
\verb?        df(Q,phi~),df(Q,phi),df(R,phi,phi~),df(Q,phi,phi~);?
\dsiline\verb?show new EQUATIONS;?
\par\noindent
The output looks as follows:
\dfoline\verb?                            ?%
\verb?                                 3    2?\npb
\dsoline\verb?  Eq. (10) reads DF(Q,phi~)=?%
\verb?( - 2*DF(k,phi,phi~,2)*I*psi*psi~ *rho  -?
\someskip
\dropout{13}
\dsoline\verb?  Eq. (9) reads DF(Q,phi)=(?
\someskip
\dsoline\rzr\verb?                                  2         2?\npb%
\dsoline\rzr\verb?2*DF(k,phi,phi~)*DF(psi,phi)*I*psi *psi~*rho ?%
\dropout{13}
\someskip
\soline{0R}\verb?  Eq. (8) reads DF(R,phi,phi~)=(?
\someskip
\dsoline\rzr\verb?                                 4     4    4?\npb%
\dsoline\rzr\verb? 4*DF(k,phi,phi~,2)*DF(k,phi)*psi *psi~ *rho ?%
\dropout{59}
\dsoline\verb?  Eq. (7) reads DF(Q,phi,phi~)=( - DF(psi,x)*psi~?%
\someskip
\dsoline\rzr\verb?                                            2?\npb%
\dsoline\rzr\verb?   + DF(psi,phi)*DF(beta~,phi~)*psi*psi~*rho ?%
\dropout{9}
\par\noindent
These \vb{EQUATIONS}
 express the content of all the field equations which have
not been satisfied so far. Their further reduction is now in order.

 Specifically, we have three equations determining derivatives of
the \vb{SCALAR} \vb{Q}. Thus they have to satisfy some
consistency conditions. Calculating \vb{2*RHS(EQ(7))%
-DR(RHS(EQ(9)),\-%
               phi\ovrl)-DR(RHS(EQ(10)),phi)} (by means of the
action \vb{EVALUATE}), one obtains identical zero. This means that
\EqN{7} does not yields a new information (and thus may be
\vb{ERASE}d). On the contrary, the condition of mutual consistency
of \EqN{9} and \EqN{10} is nontrivial. We shall see it reduces in
fact to the equation determining the \vb{SCALAR k}. The optimal rout
of its derivation is the following.

 Let us introduce the auxiliary variable \Vb{expr} which initially
is assigned with the value \vb{Q}. Then we make a series of additive
transformations of \vb{expr}. Simultaneously we shall calculate
another auxiliary variable \vb{aux} by the formula
\vb{aux=DR(expr,phi)}, replacing here the derivative of \vb{Q} by its
expression provided by \EqN{9}. The corresponding instructions
read:
\dfiline\verb?evaluate expr=Q,?
\dsiline\verb?          expr=expr -df(k,phi,phi~)?
\dsiline\verb?                     *COEFFN(NUMR(RHS(EQ(9))),df(k,phi,2,phi~))?
\dsiline\verb?                     /DENM(RHS(EQ(9))),?
\dsiline\verb?          aux=DR(expr,phi)-(LHS(EQ(9))-RHS(EQ(9))),?
\dsiline\verb?          aux;?
\dsiline\verb?evaluate  expr=expr -k*COEFFN(NUMR(aux),df(k,phi))/DENM(aux),?
\dsiline\verb?          aux=DR(expr,phi)-(LHS(EQ(9))-RHS(EQ(9))),?
\dsiline\verb?          aux;?
\dsiline\verb?evaluate  expr=expr?
\dsiline\verb?               -df(psi,x)*COEFFN(NUMR(aux),df(psi,x,phi))/DENM(aux)?
\dsiline\verb?               -beta*COEFFN(NUMR(aux),df(beta,phi))/DENM(aux),?
\siline{dQ}\verb?          aux=DR(expr,phi)-(LHS(EQ(9))-RHS(EQ(9))),?
\dsiline\verb?          aux;?
\par\noindent
The output of the last line reads:
\dfoline\verb?  The expression?
\dsoline\verb?      aux?
\dsoline\verb?                  amounts to:?
\someskip
\dsoline\verb?      psi~*xi?\npb
\dsoline\verb?     ---------                ?\rzr\verb?0.6 s.?\npb
\dsoline\verb?         2?

 In accordance with line \fn{dQ} \vb{aux} equals {\it modulo\/}
\EqN{9} the derivative of \vb{expr} with respect to \vb{phi}. This
dependence can be integrated.
Let us notice also that
\vb{expr} possesses a complex value. The instructions
\dfiline\verb?turn on the displaying of negative powers,?
\dsiline\verb?        the support of complex numbers;?%
        \footnote{this one may be shortened to \Vb{turn on DIV,COMPLEX}}
\dfiline\verb?evaluate (expr+C.C.(expr))/2,(expr-C.C.(expr))/2/i;?
\par\noindent
yield the output:
\dfoline\verb?  The expression?
\dsoline\verb?       (expr + C.C. (expr))/ 2?
\dsoline\verb?                  amounts to:?
\someskip
\dsoline\verb?      1               -1    1                 -1        1?\npb
\dsoline\verb?     ---*DF(psi,x)*psi   + ---*DF(psi~,x)*psi~   ?%
\verb?+ Q - ---*psi*beta~?\npb
\dsoline\verb?      4                     4                           4?
\someskip
\dsoline\verb?         1?\npb
\dsoline\verb?      - ---*beta*psi~            ?\rzr\verb?0.2 s.?\npb
\dsoline\verb?         4?
\someskip
\dsoline\verb?  The expression?
\dsoline\verb?       (expr - C.C. (expr))/ 2/ i?
\dsoline\verb?                  amounts to:?
\someskip
\dsoline\verb?                                   2    I               -1?\npb
\dsoline\verb?      - DF(k,phi,phi~)*psi*psi~*rho  - ---*DF(psi,x)*psi?\npb
\dsoline\verb?                                        4?

\dsoline\verb?         I                 -1    1                I?\npb
\dsoline\verb?      + ---*DF(psi~,x)*psi~   - ---*k*psi*psi~ ?%
\verb?- ---*psi*beta~?\npb
\dsoline\verb?         4                       2                4?
\someskip
\dsoline\verb?         I?\npb
\dsoline\verb?      + ---*beta*psi~            ?\rzr\verb?0.2 s.?\npb
\dsoline\verb?         4?
\par\noindent
 Thus after the integrating mentioned above, the separation of the
real and imaginary parts of the equation obtained obviously yields
explicit representation of \vb{Q} and \vb{DF(k,phi,phi\ovrl)},
respectively.

This result accomplishes the current (third) step of our calculations.

\subsection{Step 4: Determination of {\protect\tt R}}

 It was shown above that unknown \vb{Q} and the derivative
\vb{DF(k,phi,phi\ovrl)} can be explicitly expressed in the form
involving the integral of the \vb{SCALAR} function \vb{xi}
(depending on \vb{x,phi,phi\ovrl}) with respect to \vb{phi}. It is
reasonable therefore to introduce another \vb{SCALAR} (we denote it
\vb{int\und xi}) whose derivative with respect to \vb{phi} equals
\vb{xi}. Another \vb{SCALAR} independent of \vb{phi} arises as a
`constant of integration'. We denote it \vb{gam\ovrl}. \vb{gam\ovrl}
may depend on \vb{x} and \vb{phi\ovrl}. (Thus it represents the
function $\bar\gamma(x,\bar\phi)$ holomorphic with respect to
$\bar\phi$.)

Accordingly, the lines {\fn{2sc}} and {\fn{3sc}} of the declaration
specifying \vb{SCALAR} dependences now has to look as follows:
\dfiline\verb?                  psi,beta(x,phi),gam~,psi~(x,phi~),?
\dsiline\verb?                  xi,int_xi(x,phi,phi~),?
\par\noindent
while its other lines are not changed.
Besides, the items
`\vb{gam \& gam\ovrl}, \vb{int\und xi \& int\und xi\ovrl}'
are added to the description of \Vb{COMPLEX CONJUGATED} objects
which now reads:
\dfiline%
\verb? Declare COMPLEX CONJUGATED phi & phi~,al & al~,psi & psi~,?
\dsiline%
\verb?                            lam & lam~,mu & mu~,zeta & zeta~,?
\dsiline\verb?                            beta & beta~, gam & gam~,?
\dsiline\verb?                            xi & xi~, int_xi & int_xi~,?
\dsiline\verb?                            ems0 & ems0~,ems1 & ems1~;?
\par\noindent
Further, the list of \vb{ABBREVIATIONS} (see lines
{\fn{1ab}--\fn{4ab}}) is replenished with two new items:
\filine{1ex}\verb?   dd_k=-k/2/rho**2?
\szline\verb?        +i/(4 rho**2*psi*psi~)?
\szline\verb?          *(aux-C.C.(aux), WHERE?
\szline%
\verb?            aux=-df(psi,x)/psi-psi*(beta~+int_xi~+gam)),?
\szline\verb?   Q_val=(1/4)*(aux+C.C.(aux), WHERE?
\siline{5ex}%
\verb?                aux=-df(psi,x)/psi+psi*(beta~+int_xi~+gam));?
\par\noindent
which represent the values of \vb{DF(k,phi,phi\ovrl)} and \vb{Q},
respectively, derived in the preceding subsection.

We shall use below the following
\dfiline\verb?Substitutions:?
\dsiline\verb?  (1)  lam->VAL(lam);?
\dsiline\verb?  (2)  zeta->VAL(zeta);?
\dsiline\verb?  (22) zeta->VAL(zeta);?
\dsiline\verb?  (3)  mu->VAL(mu);?
\dsiline\verb?  (33) mu~->VAL(mu~);?
\dsiline\verb?  (4)  rho->VAL(rho);?
\dsiline\verb?  (5)  ems1->VAL(ems1);?
\dsiline\verb?  (55) ems1~->VAL(ems1~);?
\dsiline\verb?  (6)  ems0->VAL(ems0);?
\dsiline\verb?  (66) ems0~->VAL(ems0~);?
\dsiline\verb?  (7)  df(xi, phi~)->DR(e**(al+al~)*psi, x);?
\dsiline\verb?  (77) df(xi~, phi)->DR(e**(al+al~)*psi~, x);?
\dsiline\verb?  (8)  df(k,phi,phi~)->dd_k;?
\dsiline\verb?  (9)  df(k,phi,2,phi~)->DR(dd_k,phi);?
\dsiline\verb?  (10) df(k,phi,phi~,2)->DR(dd_k,phi~);?
\dsiline\verb?  (11)  Q->Q_val;?
\dsiline\verb?  (12)  df(Q,phi,phi~)->DR(Q_val,phi,phi~);?
\dsiline\verb?  (13)  df(Q,phi)->DR(Q_val,phi);?
\dsiline\verb?  (14)  df(Q,phi~)->DR(Q_val,phi~);?
\dsiline\verb?  (15)  df(int_xi, phi)->xi;?
\dsiline\verb?  (16)  df(int_xi~,phi~)->xi~;?
\dsiline\verb?  (17)  df(int_xi, phi,phi~)->df(xi,phi~);?
\dsiline\verb?  (18)  df(int_xi~,phi,phi~)->df(xi~,phi);?
\dsiline\verb?  (19)  xi->df(int_xi, phi);?
\dsiline\verb?  (20)  xi~->df(int_xi~, phi~);?
\dsiline\verb?end of substitutions.?
\par\noindent
Their meaning is manifest. \Rmr{Let us remind that the macro
\Vb{VAL} returns the \vb{SCALAR VALUE} of
its argument, a \vb{SCALAR} identifier.}

 The aim of the present step of calculations is to express the last
unknown \vb{SCALAR} \vb{R} through the other \vb{SCALARS} which may
be regarded in this context as known functions. More precisely, we shall
determine \vb{df(R,phi,phi\ovrl)} (essentially, the 2-dimensional
Laplacian of \vb{R}), \cf\ line {\fn{0R}}. \vb{R} is then obtained
by means of a straightforward integrating (the solving of Poisson
equation).

First of all, we deduce the equation determining \vb{R} from
the Einstein-Maxwell equations by means of the
executing appropriate substitutions of
\vb{SCALAR VALUES}. Specifically, the instructions
\dfiline\verb?obtain the SPINOR PART OF EINSTEIN-MAXWELL EQUATIONS;?
\dsiline\verb?match substitution rules (12),(17),(18) with?
\dsiline\verb?      the ABOVE EQUATIONS;?
\dsiline\verb?renew the ABOVE EQUATIONS;?
\dsiline\verb?match substitution rules (15),(16) with?
\dsiline\verb?      the ABOVE EQUATIONS;?
\dsiline\verb?renew the ABOVE EQUATIONS;?
\dsiline\verb?match substitution rules (7),(77) with?
\dsiline\verb?      the ABOVE EQUATIONS;?
\dsiline\verb?renew the ABOVE EQUATIONS;?
\dsiline\verb?match substitution rules (2),(3),(5),(6),(22),(33),(55),(66) with?
\dsiline\verb?      the ABOVE EQUATIONS;?
\dsiline\verb?renew the ABOVE EQUATIONS;?
\dsiline\verb?match substitution rules (9),(10),(13),(14) with?
\dsiline\verb?      the ABOVE EQUATIONS;?
\dsiline\verb?renew the ABOVE EQUATIONS;?
\dsiline\verb?match substitution rules (15),(16),(8),(11) with?
\dsiline\verb?      the ABOVE EQUATIONS;?
\dsiline\verb?renew the ABOVE EQUATIONS;?
\dsiline\verb?isolate the ABOVE EQUATIONS;?
\par\noindent
entails all the field equations which are not immediately satisfied
by virtue of
the relationships introduced in frames of
the current
problem specification. The result is
two \vb{EQUATIONS}. The first of them, \EqN{1}, is in fact trivial.
It is satisfied, provided the meaning of \vb{rho}
is taken into account (\cf\ the comment following lines
{\fn{0pd}--\fn{4pd}}).
Concerning another one, it may be considered just as the equation
determining \vb{R}. The instructions
\dfiline\verb?   resolve Eq. (2) w.r.t. df(R,phi,phi~);?
\dsiline\verb?   show new EQUATION;?
\par\noindent
yield in particular the following the output:
\dfoline\verb?  Eq. (3) reads DF(R,phi,phi~)=(?
\someskip
\dsoline\verb?                                               4     4    2?\npb
\dsoline\verb?                   - 8*DF(k,phi)*DF(k,phi~)*psi *psi~ *rho?
\dropout{119}
\par\noindent
({\sc r.h.s.}\ here involves neither \vb{R} nor its derivatives).
The system of field equations in the representation
described by the current problem specification
entails no other relationships which involve \vb{SCALAR R}.

Although \EqN{3} seems to be fairly involved, it is nevertheless quite
tractable. We apply the method of the `partial integrating' which
was, essentially,  already used above.
The corresponding
instructions are the following:
  \dfiline\verb? evaluate?
\siline{2R}\verb?   expr=R,?%
  \dsiline\verb?   aux=LHS(EQ(3))-RHS(EQ(3)),?
\siline{4R}\verb?   expr=expr +k**2*COEFFN(COEFFN(NUMR(aux)?
\szline\verb?                                ,df(k,phi)),df(k,phi~))?
\siline{6R}\verb?               /DENM(aux)/2,?%
\siline{7R}\verb?   aux=DR(expr,phi,phi~)-(LHS(EQ(3))-RHS(EQ(3))),?%
  \dsiline\verb?   aux=MATCHING(aux,8),?
\someskip%
 \siline{9R}\verb?   expr=expr-k*DF(psi~,x)?
\szline\verb?             *COEFFN(COEFFN(NUMR(aux),df(k,phi))?
\szline\verb?                           ,df(psi~,x,phi~))/DENM(aux)?
\szline\verb?            -k*DF(psi,x)?
\szline\verb?             *COEFFN(COEFFN(NUMR(aux),df(k,phi~))?
\siline{14R}%
\verb?                            ,df(psi,x,phi))/DENM(aux),?
\siline{15R}\verb?   aux=DR(expr,phi,phi~)-(LHS(EQ(3))-RHS(EQ(3))),?%
   \dsiline\verb?   aux=MATCHING(aux,8),?
\someskip
\siline{17R}\verb?   expr=expr-k*psi~?
\szline\verb?             *COEFFN(COEFFN(NUMR(aux),df(k,phi))?
\szline\verb?                            ,df(psi~,phi~))/DENM(aux)?
\szline\verb?            -k*psi?
\szline\verb?             *COEFFN(COEFFN(NUMR(aux),df(k,phi~))?
\siline{22R}%
\verb?                            ,df(psi,phi))/DENM(aux),?
\siline{23R}\verb?   aux=DR(expr,phi,phi~)-(LHS(EQ(3))-RHS(EQ(3))),?%
   \dsiline\verb?   aux=MATCHING(aux,8),?%
   \dsiline\verb?   aux=MATCHING(aux,17,18),?%
   \dsiline\verb?   aux=MATCHING(aux,15,16,7,77),?
\someskip
\siline{27R}\verb?   expr=expr-beta*beta~?
\szline\verb?             *COEFFN(COEFFN(NUMR(aux),df(beta,phi))?
\siline{29R}%
\verb?                            ,df(beta~,phi~))/DENM(aux),?
\siline{30R}\verb?   aux=DR(expr,phi,phi~)-(LHS(EQ(3))-RHS(EQ(3))),?%
   \dsiline\verb?   aux=MATCHING(aux,8),?%
   \dsiline\verb?   aux=MATCHING(aux,17,18),?%
   \dsiline\verb?   aux=MATCHING(aux,15,16,7,77),?%
   \dsiline\verb?   aux=MATCHING(aux,19,20),?
\someskip
\siline{35R}\verb?   expr=expr-beta*(int_xi~+gam)?
\szline\verb?             *COEFFN(COEFFN(NUMR(aux),df(beta,phi))?
\szline\verb?                           ,df(int_xi~,phi~))/DENM(aux)?
\szline\verb?            -beta~*(int_xi+gam~)?
\szline\verb?             *COEFFN(COEFFN(NUMR(aux),df(beta~,phi~))?
\siline{40R}%
\verb?                           ,df(int_xi,phi))/DENM(aux),?
\siline{41R}\verb?   aux=DR(expr,phi,phi~)-(LHS(EQ(3))-RHS(EQ(3))),?%
    \dsiline\verb?   aux=MATCHING(aux,8),?%
    \dsiline\verb?   aux=MATCHING(aux,17,18),?%
\siline{44R}\verb?   aux=MATCHING(aux,15,16,7,77);?%
\par\noindent
Their meaning is quite straightforward.
Initially, the working variable \vb{expr} is endowed with the value \vb{R}
(line {\fn{2R}}). Further it is modified by means of the adding
appropriate terms
(lines
 {\fn{4R}--\fn{6R},
 \fn{9R}--\fn{14R},
 \fn{17R}--\fn{22R},
 \fn{27R}--\fn{29R},
 \fn{35R}--\fn{40R}}) proportional to
 \vb{k**2}, \vb{k*DF(psi\ovrl,x)}, \vb{k*DF(psi,x)}, \etc{}
 The necessary coefficients are determined `automatically', utilizing the
value of the another working variable \vb{aux}. The latter is each
time equal {\it modulo\/} \EqN{3} to the second order derivative
\vb{DR(expr,phi,phi\ovrl)} of the current value of \vb{expr}, (lines
{\fn{7R}, \fn{23R}, \fn{30R}, \fn{41R}}). During these
transformations the properties (in fact, definitions) of various
variables expressed in the form of substitution rules (see above the
section of \vb{Substitutions}) are taken into account. (These are
the instructions which involve the macro \Vb{MATCHING}.) In
particular, after the execution of the instruction displayed in line
{\fn{44R}}, \vb{aux} still equals the derivative
\vb{DR(expr,phi,phi\ovrl)} {\it modulo\/} \EqN{3}. Moreover, it is
easy to see that this relationship just expresses (in a modified but
equivalent form) the essence of \EqN{3}.

 Ultimately, the instructions
\dsiline\verb?turn on the displaying of negative powers;?
\dsiline%
\verb?evaluate aux,expr factoring out R,k,df(psi,x),df(psi~,x);?
\par\noindent
entails the
following output:
\dfoline\verb?  The expression?
\dsoline\verb?      aux?
\dsoline\verb?                  amounts to:?
\someskip
\dsoline\verb?                 2    -3     -1    -2?\npb
\dsoline\verb?      - DF(psi,x) *psi  *psi~  *rho?
\someskip
\dsoline\verb?         1                          -2     -2    -2?\npb
\dsoline\verb?      - ---*DF(psi,x)*DF(psi~,x)*psi  *psi~  *rho?\npb
\dsoline\verb?         2?
\someskip
\dsoline\verb?         1               -2    -2?\npb
\dsoline\verb?      + ---*DF(psi,x)*psi  *rho  *(gam~ + int_xi)?\npb
\dsoline\verb?         4?
\someskip
\dsoline\verb?                  2    -1     -3    -2?\npb
\dsoline\verb?      - DF(psi~,x) *psi  *psi~  *rho?\npb
\someskip
\dsoline\verb?         1                 -2    -2?\npb
\dsoline\verb?      + ---*DF(psi~,x)*psi~  *rho  *(gam + int_xi~)?\npb
\dsoline\verb?         4?
\someskip
\dsoline\verb?         1                 -2     -1    -2?\npb
\dsoline\verb?      + ---*DF(psi,x,2)*psi  *psi~  *rho?\npb
\dsoline\verb?         2?
\someskip
\dsoline\verb?         1                  -1     -2    -2    1?\npb
\dsoline\verb?      + ---*DF(psi~,x,2)*psi  *psi~  *rho   - ---*xi*xi~    ?%
\rzr\verb?0.4 s.?\npb
\dsoline\verb?         2                                     4?
\someskip
\dsoline\verb?  The expression?
\dsoline\verb?      expr?
\dsoline\verb?                  amounts to:?
\someskip
\dsoline\verb?          1   2             1                   -1?\npb
\dsoline\verb?     R + ---*k *psi*psi~ - ---*k*DF(psi,x)*I*psi?\npb
\dsoline\verb?          4                 4?
\someskip
\dsoline\verb?         1                     -1         1               1?\npb
\dsoline\verb?      + ---*k*DF(psi~,x)*I*psi~   + k*I*(---*psi*beta~ + ---*psi*gam?\npb
\dsoline\verb?         4                                4               4?
\someskip
\dsoline\verb?            1                 1               1?\npb
\dsoline\verb?         + ---*psi*int_xi~ - ---*beta*psi~ - ---*gam~*psi~?\npb
\dsoline\verb?            4                 4               4?
\someskip
\dsoline\verb?            1                  1?\npb
\dsoline\verb?         - ---*psi~*int_xi) + ---*(beta*beta~ + beta*gam?\npb
\dsoline\verb?            4                  4?
\someskip
\dsoline\verb?         + beta*int_xi~ + gam~*beta~ + int_xi*beta~)        ?%
\rzr\verb?0.2 s.?
\par\noindent
In view of all said above
the flat Laplacian of \vb{expr} equals \vb{aux} which may be considered
as a known function. This relation is easily integrable.

It is evident that the result deduced accomplishes the solving of
the problem considered.

\subsection{The solution}\label{res}

 Let us summarize the above results re-casting the formulary input
data and computer output to standard mathematical notations. Namely, we
found a generic solution of Einstein--Maxwell equations which complies
with Eq.\ (\ref{raev}). Its metric is described by the expansion
(\ref{metr}) where the tetrad of 1-forms $\theta^a$ is defined as
follows
 (\cf\ \cite{Cat}, Eq.\ (27.46)):
\begin{eqnarray}
 \theta^0 &=&
\rho^{-1}\psi^{-1}\dex \phi/\sqrt{2},
\enskip\theta^1=\overline{\theta^0},
\enskip\theta^2=
\dex x/\sqrt{2},
                                                \nonumber\\
\enskip\theta^3&=&
\left(
\dex y
-2\,\im\left(k,_{\phi}\dex \phi\right)
+\left(R+Q\,y-\Ratio{1}{4}\vert\psi\vert^2 y^2\right)\dex x
\right)/\sqrt{2}
                                                \label{msol}
\end{eqnarray}
while the non-zero components of the undotted spinor of the
electromagnetic field comprise
\begin{equation}
\mbEEM{1}=\psi \mbox{\ and\ }
\mbEEM{2}=\rho\psi(-S+\psi,_{\phi}y).
                                                \label{emsol}
\end{equation}
 Here $x,y$ are the real coordinates, $\phi$ is the complex one
(whose complex conjugated
counterpart $\bar\phi$ is considered as independent
variable), the other symbols denote their functions. Specifically,
\begin{eqnarray}
\rho&=&1+\Ratio{1}{4}|\phi|^2,
                              \\
S&=&\xi+\beta,_{\phi}+2\imu\psi\,k,_{\phi}+\imu k\,\psi,_{\phi},
                              \\
R&=&D-\Ratio{1}{4}\left(|\beta|^2+k^2|\psi|^2\right)
                              \nonumber\\
&&+2\re\left(
\nu+\imu\,k\left(
\psi^{-1}\psi,_{x}-\psi
\left(
\overline{\Upsilon}+\gamma+\overline\beta\right)
-\beta\left(\overline{\Upsilon}+\gamma\right)
\right) \right),
                              \\
Q&=&\half\re\left(
\psi^{-1}\psi,_{x}-\psi\left(
\overline{\Upsilon}+\gamma+\overline\beta
\right)\right).
\end{eqnarray}
Here {\it real\/} $D=D(x,\phi,\overline\phi)$ and
$k=k(x,\phi,\overline\phi)$ satisfy the equations
\begin{eqnarray}
D,_{\phi\bar\phi}&=&
-\half\rho^{-2}|\left(\psi^{-1}),_{x}|^2-\Ratio{1}{4}|\xi|^2 \right.
                                               \nonumber\\
&&-\half\rho^{-2}\re
\left( 2\,\bar\psi^{-1}\left(\psi^{-1}\right),_{xx}
+      \left(\psi^{-1}\right),_{x}
       \left(\Upsilon+\overline\gamma\right)
\right),
                                                \nonumber
                                                \\
k,_{\phi\bar\phi}+\half\rho^{-2}k&=&
-\half\rho^{-2}|\psi|^{-2}\im\left(
\psi^{-1}\psi,_{x}+\psi\left(
\overline{\Upsilon}+\gamma+\overline\beta
\right) \right).
                                                 \label{k-eq}
\end{eqnarray}
{\it Complex\/}  functions
$\xi=\xi(x,\phi,\bar\phi)$,
$\Upsilon=\Upsilon(x,\phi,\bar\phi)$
are defined
according to the following relationships:
\begin{equation}
\xi,_{\bar\phi}=\rho^{-2}\left({\bar\psi}^{-1}\right)\!\!,_{x},
\enskip
\Upsilon,_{\phi}=\xi.
                                                 \label{endsol}
\end{equation}
The functions
$\psi=\psi(x,\phi)\ne 0$,
$\beta=\beta(x,\phi)$,
$\gamma=\gamma(x,\phi)$,
$\nu=\nu(x,\phi)$
are arbitrary. They are assumed to be holomorphic with
respect to the second (complex) argument (and smooth with respect to
the first one).

It is worth mentioning that although the description of the solution
presented involves some unsolved equations, all they are solved
by means of subsequent quadratures. Indeed, there are three kinds
of equations which, dropping out irrelevant complicating details,
look as follows:
\begin{eqnarray}
&(*)&{\partial f(z,\bar z)
\over
 \partial z}=g(z,\bar z),\qquad
(**)\quad
{\partial^2 f(z,\bar z)
\over
 \partial z\partial \bar z}=g(z,\bar z),
                              \nonumber\\
&(*\!*\!*)\quad&(1+z\bar z)^2
{\partial^2 f(z,\bar z)
\over
 \partial z\partial \bar z}+2\,f
=g(z,\bar z).
\nonumber
\end{eqnarray}
Here $g=g(z_1, z_2)$ denotes some known function holomorphic with
respect to the {\em both\/} arguments, \ie\ admitting a holomorphic
extension to some neighborhood of the `shell'
$\{z_1={\bar z}_2\}\subset \Cset^2$. Local solutions $f$ of
these equations share the latter property. They can
be represented in explicit form in terms of quadratures as follows:
 \begin{eqnarray}
&\ma&
f=z\int\limits^1_0 \dex  t\, g(t\,z,\bar z),\qquad
\maa\quad
f=z\bar z\int\limits^1_0\!\!\int\limits^1_0
 \dex  t_1\, \dex  t_2\, g(t_1z,t_2\bar z),
                              \nonumber\\
&\maaa\quad&
f=z\bar z\int\limits^1_0\!\!\int\limits^1_0
\dex  t_1\, \dex  t_2\,
\left(
1-
{2 z \bar z (1-t_1)(1-t_2)
\over
(1+z\bar z)(1+z\bar z t_1 t_2)}
\right)
g(t_1 z,t_2\bar z)
                              \nonumber\\
&&
\hphantom{f=}
+\re\left(
{\dex  j(z)\over \dex  z}
         -
         {2\bar z j\left(z\right)
\over
          1+z\bar z}
\right),
\nonumber
\end{eqnarray}
 the last formula (local general solution of inhomogeneous Helmholz
equation on a sphere) being of certain independent interest. Here
$j=j(z)$ is an arbitrary holomorphic function playing role of
`integration constant'. It is worth noting that the second term in
$\maaa$ represents itself the general local solution of the
homogeneous version ($g=0$) of Eq.\ $(*\!*\!*)$ and thus $\maaa$ is
its general local solution in non-homogeneous case. In the formulae
$\ma$, $\maa$ the `integration constants' similar to $j(z)$ are
dropped out since they had been explicitly introduced in the above Eqs.\
(\ref{msol}--\ref{endsol}) describing the solution of the field
equations. This is just the origin of the arising of the arbitrary
functions $\nu, \gamma, \beta$. On the contrary, the arbitrary
function which arises when integrating Eq.\ (\ref{k-eq}) (equivalent
to $(*\!*\!*)$) with respect to $k$ was not explicitly introduced,
being implicitly involved in (the value of) $k$. It is the fifth
arbitrary function (the fourth one is $\psi$).

Thus, in total, the electrovac solution obtained involves
{\em five arbitrary functions of two variables\/}
holomorphic with respect to one of them.

Concerning the aspect of the physical interpretation of the
field configuration described by equations (\ref{msol}--\ref{endsol}), one
should take into account that a huge `amount' of degrees of freedom
the family of solutions possesses makes fairly difficult, if not
impossible, to reveal and formulate it in a full generality.
Nevertheless one may gain some insight into the problem
analyzing suitable particular cases (subfamilies) of the field
configurations described by the above equations. That way, a useful simple
example is provided by the following their representative (\cf\
\cite{Cat}, Eq.\ (27.54)):
\begin{eqnarray}
\mt&=&
\dex x \sot
\left(
\dex y
+\left(\re\nu(x,\phi)
-\vert\psi(\phi)\vert^2 y^2\right)\dex x
\right)
+
{\dex \phi \sot \dex \bar\phi
\over
|\psi(\phi)|^2 \left(1+|\phi|^2\right)^2
},
                                        \nonumber\\
\omega&=&
-\dex (\psi y)
+{
\dex  \phi\wedge\dex\bar\phi \over
\bar\psi \left(1+|\phi|^2\right)^2
}.
                                                \nonumber
\end{eqnarray}
 (Here, with regard to Eqs.\ (\ref{msol}--\ref{endsol}), a minor
modification of the gauge and some obvious elementary re-definitions
were carried out). The arbitrary functions $\psi(\phi), \nu(x,\phi)$
are holomorphic with respect to $\phi$.

As opposed to the general case,
 the meaning of the latter field configuration is immediately manifest:
it represents
the generalized plane wave spread against the Bertotti--Robinson
space-time \cite{BR} describing itself non-null homogeneous
static electromagnetic field. Indeed, the Bertotti-Robinson field
configuration arises in the case $\nu=0, \psi=\mbox{constant}\ne 0$.
On the other hand an appropriate limiting procedure corresponding
to the nullifying the static electric (magnetic) field yields the
{\em pp}-wave metric.

\section{Summary and discussion}

 Summarizing, the paper presented exhibits an example demonstrating the
application of the specialized computer algebra system \Grg\ to the
problem of the searching for solutions to the coupled Maxwell and
Einstein--Maxwell equations. We investigated their general class
complying ansatz (\ref{raev}). The latter equation imposes
certain restrictions on the irreducible curvature spinors
which give evidence (basing first of all on important
particular cases) for the associating the configurations under
consideration with radiation processes.

Reassigning all the routine calculations to \Grg, we found a
generic solution of the problem which turns out to involve five
arbitrary functions of two arguments (see subsection \ref{res}).
In agreement, with \apriori\ expectations its particular case
can be interpreted as {\em pp}--wave spread against a background
Bertotti-Robinson space-time, see Refs.~\cite{BR}, \cite{Cat}.
This circumstance partially justifies the
working term `radiative electrovacs'
 (perhaps too wide in the context considered)
which was used above for the referring to the class of
space-times investigated.

 The complete problem specification (\Grg\ {\em input code}) which
encodes the formulae (\ref{msol}--\ref{endsol}) from subsection
\ref{res} and represents the solution found is displayed in Appendix
A. Appendix B displays the corresponding \Grg\ output which proves
that we actually deal with a solution of the relevant field
equations%
  \footnote{For the sake
            to provide a performance scale for the timing labels shown in the
            output protocol
            (as well as for incidental ones occurred above in the fragments of
            output listings reproduced), we note that the IBM/PC compatible
            computer with AMD386DX 40 MHz CPU and 8 Mb of RAM was
            used. A fiducial point characterizing its (fairly moderate)
            performance rate is provided by the time of the expanding of
            $(a+b+c)^{100}$ by \Red\ (without output). The corresponding
            script reads `\vb{showtime\$}\vb{(A+B+C)**100}\vb{\$showtime;}'.
            It reports about 87 seconds of CPU labour.
}.

 As it has been mentioned, the prevailing aim pursued in the present
work is the characterizing of \Grg\ system from viewpoint of its
suitability for a practical application. Additionally, we
simultaneously attempted, mostly after indirect fashion, to exhibit
possible styles of its usage and to demonstrate some of its
capabilities. In particular, this is one of the purposes
the inclusion of section
\ref{grgappl} which discusses the processing of the problem
in the complete form.

 It is instructive to mention that the treatment of the problem
considered in subsections \ref{s2},\ref{s3} began
immediately with consideration of a {\em workable \mbox{\rm\Grg}
program\/} (we called it problem specification, see subsection
\ref{start}). In particular we did not need, essentially,  to
preliminarily learn \Grg\ input language and data
stuctures to a notable extent, restricting ourselves to several
remarks. At the same time it should be mentioned that, in principle,
\Grg\ is a fairly complex system whose complete description occupies
hundreds of pages. One can estimate therefore a prominent clarity of
the organization of \Grg\ language which enables one to comprehend
the essence of \Grg\ input without becoming absorbed into
specific programming-related issues%
  \footnote{The developing of new programs requires more detailed
           knowledge of course.}.
 The content of Appendix A, where an integrated problem specification
is exhibited, provides a nice demonstration of the latter circumstance.

 To be more specific, one may distinguish, in principle, the two kinds
of \Grg\ code. The first one mostly follows a traditional imperative
style of wellknown programming languages. Typical examples of
such a sort scripts are displayed in lines {\fn{0hi}--\fn{9hi}},
{\fn{1ex}--\fn{5ex}}, \etc{} Generally speaking, the instructions
contained in these fragments of \Grg\ script realize, in a sense, `low
level' mathematical calculations handling symbolic formulae.
 Their evident characteristic property is, as usually, a high degree
of a detailed petty control by a user depending on intermediate
results required. It should be noted however that \Grg\ provides a
rather limited set of facilities intended for such a `lower level'
algebraic programming. For example, no loops, conditional and
branching operators, subroutines, \etc, are supported. The point is
that, though it might seem surprising, they are superfluous here in
fact.

 A foundation (and, hopefully, a source of potential advantages)
of \Grg\ lives in a distinct area. Essentially, in \Grg\ the control
over majority of calculations is based mostly on {\it actions\/} of
a {\em maximally high level\/}, approaching the one which might be
attributed to the relationships and notions characteristic of the
application field (the geometry and the field theory) itself. As examples
of the corresponding programming (a code of the `second' kind) one
could get the instructions displayed in lines {\fn{2k}--\fn{4k}} or
{\fn{5k},\fn{6k},\fn{7k}}. The series of instructions displayed below
(having no relation to the problem considered in the present work)
also serves an illustration of such a `super-high level' coding:
\dfiline\verb?  find UNDOTTED WEYL SPINOR and compare it with sample;?
\dsiline\verb?  classify the ABOVE STUFF;?
\dsiline\verb?  calculate factorizing denominators?%
\verb? UNDOTTED WEYL INVARIANTS?
\dsiline\verb?            and write them to a disk file;?
\par\noindent

Illuminating the grounds of the approach utilized,
 one of the cornerstones of a `knowledge' implemented in \Grg\ and
concerning the mathematical and physical theoretical issues is the
collection of so called {\em data objects\/} which model the basic
notions originated from the geometry and the field theory. In the
present work, we dealt with a fairly moderate part of them which
included \vb{TETRAD}, \vb{UNDOTTED CONNECTION}, \vb{UNDOTTED EM
SPINOR}, and some others. A full family of data objects implemented
so far is rather numerous (although still not exhaustive) and
also includes such entities as, for example,
\vb{the DATA SAVED} (to a disk storage),
\vb{ALL the TRACES OF ENERGY-MOMENTA}
(known in the current point of calculation),
\vb{COVARIANT DIFFERENTIALS OF WEYL 2-FORMS},
and even
\vb{DUPLICATED CHARGED DIRAC DERIVATIVE OF UNDOTTED DIRAC
PHI-SPINOR}.
 These are {\em export names\/} of data objects which are
used for the referring to them in instructions. (There is also
possibilities of the access to separate components, if any, of a data object%
. The corresponding
examples can be found in lines {\fn{44}, \fn{45}, \fn{50}, \fn{51}}
and others.) In order to determine the {\em value\/} of some data
object one should apply one of the {\em actions\/} \Vb{ELICIT} (from
a \vb{DATA} included in the problem specification), \Vb{CALCULATE}
(from the other data objects), \Vb{FIND} (\ie\ partially \Vb{ELICIT}
and partially \Vb{CALCULATE}), \Vb{OBTAIN} (used for equations).
Depending on the current state of environment, the realization of
such actions can be rather sophisticated invoking, in particular,
(the models of) various relationships originated from the physical
theory and geometry. Further, given the value of a data object, one
may use for its transformation various {\em substitution rules\/}
applying them
either globally (by means of the actions \vb{EXCITE}
or \vb{LET}) or locally (action \vb{MATCH}). The substitutions with
the help of \Red' package \Com\ are realised by the action \vb{MATCH
COMPACTIFYING}.
   The \Red'
polynomial factorizer is invoked by means of the action
 \Vb{RENEW FACTORIZING}. Alternatively, one may
 \Vb{RENEW FACTORIZING NUMERATORS} or
 \Vb{RENEW FACTORIZING DENOMINATORS}.
The \Red' routine performing decomposition of rational
functions into simple fractions is invoked by means of the
instruction `\vb{FRACTIONATE} \ntt{data}{object}
\vb{WITH RESPECT TO}%
        \footnote{These three `keywords' may be shortened to \Vb{w.r.t.}.}
\nt{kernel}%
 \footnote{This term closely corresponds to the notion of a {\it kernel\/}
          introduced and supported in \Red.}'
(or `\vb{FRACTIONATE CONDENSING}\dots' which additionally re-arranges
simple denominators into factors-\vb{SCALARS}), \etc

Generally speaking, we have listed major part of actions intended
for the `active' handling of data objects%
%
.
 One sees that, in principle, the controlling facilities
supported by \Grg\ are in no way intricate while they reveal no
potential limitations on the complexity of the
calculation processes underlaid.
 It may be stated that \Grg\ successively follows a tendency to
ensure a maximal clarity and naturalness of the applied code.

It is also worth noting that \Grg\ usually does not advertise about the
ways of the realizing of the instructions
performed. The listing displayed in Appendix B is an instructive
illustration of the latter circumstance%
\footnote{A copy of output directed to a monitor screen is there
presented. \grg\ additionally stores the output listing
which contains somewhat more detailed information to a disk file.}.
Indeed, one finds no formulae in the output issued by the
instructions executed. Nevertheless the main result --- the
confirmation of the satisfaction of the relevant equations --- is
clearly exhibited. (If one wants to be informed in more details, an
additional instruction to \Vb{type ALL KNOWN} would give rise to a
lot of formulae.)

Similarly, as
it was mentioned, the basic mathematical formalism which is
used for majority of calculations discussed in the present work is
the calculus of exterior forms%
   \footnote{\grg\ uses its own implementation independent of the
            \Red{}' sub-package \Exc{}.}%
. However, this circumstance nowhere explicitely manifests itself
(except perhaps of instruction \fn{1exf} and
lines \fn{10g}, \fn{20g} of output).
This may be interpreted as a particular manifestation of the intention
to release a user from superfluous details.

 At the same time the implementation of exterior calculus (together
with the standard tensor methods) constitutes itself a foundation of
the library of next level routines realizing geometrical and
physical theoretical relationships. These two interconnected
underlaid `strata' of \Grg\ system are relatively independent from
their `manager' --- the interpreter. On the other hand the narrow
application field of programming system
is determined mostly by the library of applied
routines (and, to a less extent, by the basic mathematical tools
implemented). Thus, having substituted another applied library
(which should support an appropriate collection of data objects), one
obtains the system with another application field which is
controlled, essentially, in the same way. Thus the approach to the
integrating of the constituents of \Grg\ outlined possesses a
flexibility which makes \Grg\ a valuable subject of a practical
interest worthing a further working up.

Resuming, it may be stated that \Grg\ system can be successfully
applied for practical calculations in the field of gravitation
theory, demonstrating a high efficiency and excellent convenience in
an application. It can be also estimated as a promising base for the
further development of the efficient tools for doing computer
analysis of a wide scope of problems in the field of theoretical
physics.

\bigskip
\appendix{\large\bf Appendix A}\label{a1}

Here an example of the problem specification (input code processed by \Grg) is
displayed. It checks the fulfillment of the relevant field
equations by the solution described in section \ref{res}.

\noindent
\strut\hfill\parbox[t]{0.3\textwidth}{\strut\hrulefill\null}\hfill\null
\dfline\verb?Problem Electrovac Metric.?
\someskip
\sline\verb?Data:?
\someskip
\sline\verb? declare COORDINATES x,y,phi,phi~;?
\sline\verb? declare SCALARS psi,beta,gam,nu(x,phi),?
\sline\verb?                 rho(phi,phi~),?
\sline\verb?                 k,xi,int_xi,int_del(x,phi,phi~);?
\someskip
\sline\verb? declare REAL x,y,k,rho,del,int_del;?
\sline\verb? declare COMPLEX CONJUGATED phi & phi~,?
\sline\verb?          psi & psi~,beta & beta~,gam & gam~, nu & nu~,?
\sline\verb?          xi & xi~,int_xi & int_xi~;?
\someskip
\sline\verb? SCALAR VALUE follows: rho=1+phi*phi~/4;?
\someskip
\sline\verb? ABBREVIATIONS comprise?
\sline\verb?        emS= xi +df(beta,phi) +2i*psi*df(k,phi) +i*k*df(psi,phi),?
\sline\verb?        dd_k=-k/rho**2/2?
\sline\verb?             +1/(4 i*rho**2*psi*psi~)?
\sline\verb?               *(aux-C.C.(aux), WHERE?
\sline\verb?                 aux=df(psi,x)/psi+psi*(beta~+int_xi~+gam)),?
\sline\verb?        Q=(1/4)*(aux+C.C.(aux), WHERE?
\sline\verb?                 aux=-df(psi,x)/psi+psi*(beta~+int_xi~+gam)),?
\sline\verb?        R=int_del?
\sline\verb?            -(beta*beta~ +k**2*psi*psi~)/4?
\sline\verb?            +(aux+C.C.(aux), WHERE?
\sline\verb?              aux=nu +i*k*(DF(psi,x)/psi -psi*(beta~+int_xi~+gam))?
\sline\verb?                  -beta*(int_xi~+gam))/4,?
\sline\verb?        del=-DR(1/psi,x)*DR(1/psi~,x)/rho**2/2 -xi*xi~/4?
\sline\verb?            -(aux+C.C.(aux), WHERE?
\sline\verb?              aux=2 DR(1/psi,x,2)/psi~?
\sline\verb?                  +DR(1/psi,x)*(int_xi+gam~))/rho**2/4;?
\sline\verb? TETRAD comprises?
\sline\verb?        component|0=d phi/psi/rho/sqrt(2),?
\sline\verb?        component|1=C.C.(component|0),?
\sline\verb?        component|2=d x/sqrt(2),?
\sline\verb?        component|3=(d y -2 IM(df(k,phi) d phi)?
\sline\verb?                     +(R +Q*y -y**2*psi*psi~/4) d x)/sqrt(2);?
\someskip
\sline\verb? UNDOTTED EM SPINOR comprises?
\sline\verb?        component|1=psi,?
\sline\verb?        component|2=rho*psi*(-emS +y*df(psi,phi));?
\someskip
\sline\verb? DOTTED EM SPINOR is HERMITEAN CONJUGATED to UNDOTTED EM SPINOR;?
\someskip
\sline\verb?end of data.?
\someskip
\sline\verb?Substitutions:?
\someskip
\sline\verb? (4)   rho -> VAL(rho);?
\sline\verb? (7)   df(xi,phi~) -> DR(1/psi~,x)/rho**2;?
\sline\verb? (77)  df(xi~,phi) -> DR(1/psi, x)/rho**2;?
\sline\verb? (8)   df(k,phi,phi~) -> dd_k;?
\sline\verb? (9)   df(k,phi,2,phi~) -> DR(dd_k,phi);?
\sline\verb? (10)  df(k,phi,phi~,2) -> DR(dd_k,phi~);?
\sline\verb? (15)  df(int_xi , phi) -> xi;?
\sline\verb? (16)  df(int_xi~,phi~) -> xi~;?
\sline\verb? (17)  df(int_xi, phi,phi~) -> df(xi,phi~);?
\sline\verb? (18)  df(int_xi~,phi,phi~) -> df(xi~,phi);?
\sline\verb? (19)  df(int_xi, phi,2) -> df(xi,phi);?
\sline\verb? (100) df(int_del,phi,phi~) -> del;?
\someskip
\sline\verb?end of substitutions.?
\someskip
\sline\verb?Instructions:?
\someskip
\sline\verb?  obtain UNDOTTED MAXWELL EQUATIONS;?
\sline\verb?  match substitution rule (7) with the ABOVE EQUATIONS;?
\sline\verb?  obtain EINSTEIN-MAXWELL EQUATIONS;?
\sline\verb?  match substitution rule (4)?
\sline\verb?         with the SCALAR PART of EINSTEIN-MAXWELL EQUATIONS;?
\sline\verb?  match substitution rules (9),(10),(17),(18)?
\sline\verb?         with the SPINOR PART of EINSTEIN-MAXWELL EQUATIONS;?
\sline\verb?  renew  the ABOVE EQUATIONS;?
\sline\verb?  match substitution rules (7),(77),(8),(15),(16) with?
\sline\verb?         the ABOVE EQUATIONS;?
\sline\verb?  renew the ABOVE EQUATIONS;?
\sline\verb?  match substitution rules (4),(100) with?
\sline\verb?         the ABOVE EQUATIONS;?
\sline\verb?  renew and type ALL the EQUATIONS;?
\someskip
\sline\verb?quit;?
\sline\verb?end of instructions.?
\sline\verb?  >>slang<<?
\sline\verb?Run!?

\bigskip
\appendix{\large\bf Appendix B}

The output of the processing by \Grg\
the code displayed
in Appendix
A is given. (Some initialization messages including
the copy of the input script are dropped out.).\\
\strut\hfill\parbox[t]{0.6\textwidth}{\strut\dotfill\null}\hfill\null
\dfline\verb?Total time spent amounts to 4.5 seconds.?
\someskip
\sline\verb?            -----  Processing of the problem `Electrovac Metric' -----?
\someskip
\sline\verb?COORDINATES are listed below:?
\someskip
\sline\verb?    x, y, phi, phi~?
\someskip
\sline\verb?SCALARS dependences are shown below:?
\someskip
\sline\verb?    nu:  (x,phi)?
\sline\verb?    gam:  (x,phi)?
\sline\verb?    beta:  (x,phi)?
\sline\verb?    psi:  (x,phi)?
\sline\verb?    rho:  (phi,phi~)?
\sline\verb?    int_del:  (x,phi,phi~)?
\sline\verb?    int_xi:  (x,phi,phi~)?
\sline\verb?    xi:  (x,phi,phi~)?
\sline\verb?    k:  (x,phi,phi~)?
\sline\verb?    nu~:  (x,phi~)       (is added as C.C. of the SCALAR nu)?
\sline\verb?    gam~:  (x,phi~)      (is added as C.C. of the SCALAR gam)?
\sline\verb?    beta~:  (x,phi~)     (is added as C.C. of the SCALAR beta)?
\sline\verb?    psi~:  (x,phi~)      (is added as C.C. of the SCALAR psi)?
\sline\verb?    int_xi~:  (x,phi,phi~)?
\sline\verb?                         (is added as C.C. of the SCALAR int_xi)?
\sline\verb?    xi~:  (x,phi,phi~)   (is added as C.C. of the SCALAR xi)?
\someskip
\sline\verb?No unrealized abbreviations have been specified...?
\someskip
\sline\verb?Abbreviations processed are listed below:?
\someskip
\sline\verb?    emS, dd_k, Q, R, del?
\someskip
\sline\verb?SCALAR VALUE is shown below:?
\someskip
\sline\verb?           phi*phi~ + 4?
\sline\verb?    rho = --------------.?
\sline\verb?                4?
\someskip
\sline\verb?No unrealized form abbreviations have been specified...?
\someskip
\sline\verb?No form abbreviations have been specified...?
\someskip
\sline\verb?Total time spent amounts to 6.1 seconds.?
\someskip
\sline\verb?                ** The instructions given will be executed now **?
\someskip
\sline\verb?==> obtain UNDOTTED MAXWELL EQUATIONS?
\someskip
\sline\verb?...UNDOTTED MAXWELL EQUATIONS have been obtained               ?\rzr\verb?0.6 s.?
\someskip
\sline\verb?Total time spent amounts to 6.7 seconds.?
\someskip
\sline\verb?==> match substitution rule (7) with the ABOVE EQUATIONS?
\someskip
\sline\verb?UNDOTTED MAXWELL EQUATIONS have been processed                  ?\rzr\verb?0.1 s.?
\someskip
\sline\verb?==> obtain EINSTEIN - MAXWELL EQUATIONS?
\someskip
\sline\verb?...SPINOR PART OF EINSTEIN - MAXWELL EQUATIONS has been obtained?\rzr\verb?6.4 s.?
\someskip
\sline\verb?...SCALAR PART OF EINSTEIN - MAXWELL EQUATIONS has been obtained?\rzr\verb?6.7 s.?
\someskip
\sline\verb?Total time spent amounts to 13.4 seconds.?
\someskip
\sline\verb?==> match substitution rule (4) with the SCALAR PART of EINSTEIN - MAXWELL?
\sline\verb?    EQUATIONS?
\someskip
\sline\verb?SCALAR PART OF EINSTEIN - MAXWELL EQUATIONS has been processed?\rzr\verb?0.2 s.?
\someskip
\sline\verb?==> match substitution rules (9), (10), (17), (18) with the SPINOR PART of?
\sline\verb?    EINSTEIN - MAXWELL EQUATIONS?
\someskip
\sline\verb?SPINOR PART OF EINSTEIN - MAXWELL EQUATIONS has been processed?\rzr\verb?2.6 s.?
\someskip
\sline\verb?==> renew the ABOVE EQUATIONS?
\someskip
\sline\verb?SPINOR PART OF EINSTEIN - MAXWELL EQUATIONS has been renewed  ?\rzr\verb?3.3 s.?
\someskip
\sline\verb?==> match substitution rules (7), (77), (8), (15), (16) with the ABOVE?
\sline\verb?    EQUATIONS?
\someskip
\sline\verb?SPINOR PART OF EINSTEIN - MAXWELL EQUATIONS has been processed?\rzr\verb?5.7 s.?
\someskip
\sline\verb?==> renew the ABOVE EQUATIONS?
\someskip
\sline\verb?SPINOR PART OF EINSTEIN - MAXWELL EQUATIONS has been renewed  ?\rzr\verb?5.9 s.?
\someskip
\sline\verb?==> match substitution rules (4), (100) with the ABOVE EQUATIONS?
\someskip
\sline\verb?SPINOR PART OF EINSTEIN - MAXWELL EQUATIONS has been processed?\rzr\verb?6.4 s.?
\someskip
\sline\verb?==> renew and type ALL the EQUATIONS?
\someskip
\sline\verb?--> RENEW ALL the EQUATIONS?
\someskip
\sline\verb?SPINOR PART OF EINSTEIN - MAXWELL EQUATIONS has been renewed?\rzr\verb?6.5 s.?
\someskip
\sline\verb?SCALAR PART OF EINSTEIN - MAXWELL EQUATIONS has been renewed?\rzr\verb?6.7 s.?
\someskip
\sline\verb?UNDOTTED MAXWELL EQUATIONS have been renewed                ?\rzr\verb?6.7 s.?
\someskip
\sline\verb?--> TYPE ALL the EQUATIONS?
\someskip
\sline\verb?SPINOR PART OF EINSTEIN - MAXWELL EQUATIONS?
\someskip
\sline\verb?    is satisfied?
\someskip
\sline\verb?SCALAR PART OF EINSTEIN - MAXWELL EQUATIONS?
\someskip
\sline\verb?    is satisfied?
\someskip
\sline\verb?UNDOTTED MAXWELL EQUATIONS?
\someskip
\sline\verb?    are satisfied?
\someskip
\sline\verb?==> quit?
\someskip
\sline\verb?Total time spent amounts to 20.3 seconds, garbage collection consumed?
\sline\verb?137 ms.?
\someskip
\sline\verb? The DISK file for the copying is being closed... ...done?
\someskip
\sline\verb?Quitting?

\end{document}